\documentclass[a4paper,fleqn,usenatbib]{mnras}
\usepackage{subfig}
\usepackage{url}
\usepackage{graphics,graphicx}
\usepackage{caption}
\usepackage{floatflt}
\usepackage{natbib}
\usepackage{verbatim}
\usepackage{longtable}
\usepackage{multirow,bigdelim,array,pdflscape,booktabs}

\newcommand{\gtsim}{\mbox{$\,\stackrel{>}{_\sim}$\,}}
\newcommand{\ltsim}{\mbox{$\,\stackrel{<}{_\sim}\,$}}

\def\mnras{{MNRAS}}
\def\apj{{ApJ}}

\def\aap{{A\&A}}
\def\apjl{{ApJL}}
\def\apjs{{ApJS}}

\def\pasp{{PASP}}
\def\ssr{{Sp. Sci. Rev.}}

\def\mcg6{{MCG-6-30-15}}
\newcommand\swift{{\it Swift}}

\def\ecs{ergs cm$^{-2}$ s$^{-1}$~}
\def\msun{$M_{\odot}$}

\def\me{{$\dot{m}_{E}$}}

\def\ltsim{\mathrel{\hbox{\rlap{\hbox{\lower4pt\hbox{$\sim$}}}\hbox{$<$}}}}
\def\gtsim{\mathrel{\hbox{\rlap{\hbox{\lower4pt\hbox{$\sim$}}}\hbox{$>$}}}}

\begin{document}

\title[Swift X-ray/UV/optical variability of NGC~4593]{X-ray/UV/optical variability of NGC~4593 with Swift: Reprocessing of X-rays by an extended reprocessor.}

\author[M$\rm^{c}$Hardy, I.M.]
{ I M M$\rm^{c}$Hardy$^{1}$, S D Connolly$^{1}$, K Horne$^{2}$ 
E M Cackett$^{3}$, J Gelbord$^{4}$, B M Peterson$^{5,6}$, 
\newauthor  M Pahari$^{7}$, N Gehrels$^{8}$, 
R. Edelson$^{9}$, M Goad$^{10}$, P Lira$^{11}$, P Arevalo$^{12}$, R D Baldi$^{1}$, 
\and N. Brandt$^{13}$, E. Breedt$^{14}$,
H Chand$^{15}$, G. Dewangan$^{7}$, C. Done$^{16}$, M Elvis$^{17}$, \newauthor 
D Emmanoulopoulos$^{1}$, M M Fausnaugh$^{5}$,
S Kaspi$^{18}$, C S Kochanek$^{5}$,  K Korista$^{19}$,\newauthor
I E Papadakis$^{20}$,  A R Rao$^{21}$,  
P Uttley$^{22}$, M Vestergaard$^{23}$, M J Ward$^{17}$
\\
\\
$^{1}$ Department of Physics and Astronomy, The University, Southampton
SO17 1BJ \\
$^{2}$ SUPA Physics and Astronomy, University of St.Andrews, North Haugh, KY16~9SS, Scotland, UK \\
$^{3}$ Department of Physics and Astronomy, Wayne State University, 666 W. Hancock Street, Detroit, MI 48201, USA \\
$^{4}$ Spectral Sciences Inc, 4 Fourth Avenue, Burlington, MA 01803 USA\\
$^{5}$ Department of Astronomy, The Ohio State University, 140 W 18th Avenue, Columbus, OH 43210, USA\\
$^{6}$ Space Telescope Science Institute, 3700 San Martin Drive, Baltimore, MD 21218, USA\\
$^{7}$ Inter-University Centre for Astronomy and Astrophysics, Pune, 411007, India\\
$^{8}$ Astrophysics Science Division, NASA Goddard Space Flight Center, Greenbelt, MD 20771, USA \\
$^{9}$ Department of Astronomy, University of Maryland, College Park, MD 20742-2421, USA\\
$^{10}$ University of Leicester, Department of Physics and Astronomy, Leicester, LE1 7RH  \\
$^{11}$ Departamento de Astronomia, Universidad de Chile, Camino del Observatorio 1515, Santiago, Chile\\
$^{12}$ Instituto de Física y Astronomía, Facultad de Ciencias, Universidad de Valparaíso, Gran Bretana N 1111, Playa Ancha, Valparaíso, Chile\\
$^{13}$ Department of Astronomy and Astrophysics, Pennsylvania State University, 525 Davey Laboratory, University Park, PA 16802, USA\\
$^{14}$ Institute of Astronomy, University of Cambridge, Madingley Road, Cambridge CB3 0HA, UK\\
$^{15}$ Aryabhatta Research Institute of Observational Sciences (ARIES), Manora Peak, Nainital, 263002 \\
$^{16}$ Department of Physics, University of Durham, South Road, Durham DH1 3LE \\
$^{17}$ Harvard-Smithsonian Center for Astrophysics, Cambridge, MA 02138\\
$^{18}$ School of Physics and Astronomy, Raymond and Beverly Sackler Faculty of Exact Sciences, Tel Aviv University, Tel Aviv 69978, Israel \\
$^{19}$ Department of Physics, Western Michigan University, 1120 Everett Tower, Kalamazoo, MI 49008-5252, USA\\
$^{20}$ Department of Physics and Institute of Theoretical and Comp utational Physics, University of Crete, 71003, Heraklion, Greece \\
$^{21}$ Department of Astrophysics and Astronomy, Tata Institute of Fundamental Research, Mumbai 40005, India\\
$^{22}$ Astronomical Institute “Anton Pannekoek,”University of Amsterdam, Postbus 94249, NL-1090 GE Amsterdam, The Netherlands\\
$^{23}$ Dark Cosmology Centre, Niels Bohr Institute, University of Copenhagen, Juliane Maries Vej 30, DK-2100 Copenhagen, Denmark
}
\maketitle
\begin{abstract}
We report the results of intensive X-ray, UV and optical monitoring of the Seyfert 1 galaxy NGC~4593 with Swift.  There is no intrinsic flux-related spectral change in the the variable components in any band with small apparent variations due only to contamination by a second constant component, possibly a (hard) reflection component in the X-rays and the (red) host galaxy in the UV/optical bands.
Relative to the shortest wavelength band, UVW2, the lags of the other UV and optical bands are mostly in agreement with the predictions of reprocessing of high energy emission from an accretion disc. The U-band lag is, however, far larger than expected, almost certainly because of reprocessed Balmer continuum emission from the more distant broad line region gas.
The UVW2 band is well correlated with the X-rays but lags by $\sim6 \times$ more than expected if the UVW2 results from reprocessing of X-rays on the accretion disc. However, if the lightcurves are filtered to remove variations on timescales $>5$d, the lag approaches the expectation from disc reprocessing.
MEMEcho analysis shows that direct X-rays can be the driver of most of the variations in the UV/optical bands as long as the response functions for those bands all have long tails (up to 10d) in addition to a strong peak (from disc reprocessing) at short lag ($<1$~d). We interpret the tails as due to reprocessing from the surrounding gas.
Comparison of X-ray to UVW2  and UVW2 to V-band lags  for 4  AGN, including NGC~4593, shows that all have UVW2 to V-band lags which exceed the expectations from disc resprocessing by $\ltsim 2$.
However the X-ray to UVW2 lags are, mostly, in greater excess from the expectations from disc reprocessing and differ between AGN. The largest excess is in NGC~4151. Absorption and scattering may be affecting X-ray to UV lags.
\end{abstract}

\section{Introduction}
\label{sec:intro}

The origin of the UV and optical variability in AGN, and its relationship to the X-ray variability, are questions of major relevance to understanding the central structures of AGN.
One possible explanation of UV/optical variability is that
variations in the thermal emission from the accretion disc are caused by fluctuations in
the inward accretion flow \citep{arevalouttley06}. A second possibility is
that X-ray emission from the central corona or very hard UV emission
from the very inner edge of the accretion disc illuminates the outer disc,
heating it up and causing it to re-radiate \citep{haardt91}.
For both possibilities, the disc has the same basic temperature
structure, hotter (dominating the UV emission) near the center and
cooler (optical) further out.

The time lag between the high energy emission and the re-radiated lower energy UV/optical
emission gives us the distance between these two
emission regions. Therefore, by measuring the lags between the high energy emission
and a number of UV/optical bands we can map out
the temperature structure of the disc. This technique is known
as `Reverberation mapping' \citep[RM:][]{blandford_mckee82} and has been
used to map regions too small to be resolved by direct imaging, eg AGN broad
line regions \citep[BLR,][]{peterson14}.

The model of a smooth, optically thick, geometrically thin,
efficiently radiating accretion disc was first derived by \citet[SS]{shakura73} 
and has been our basic disc model for over 40 years. In this model the release of
gravitational potential energy from accreting material leads to a
temperature profile (in physical units) of
$T(R) \propto R^{-3/4} (M \dot{m})^{1/4}$. Incident high energy emission
will enhance the existing thermal emission (slightly altering
the disc temperature profile). We thus expect a wavelength ($\lambda$)
dependent lag, $\tau$, between the incident high energy, and re-radiated UV/optical emission, of
$\tau = R/c \propto (M^{2}\dot{m}_{E})^{1/3} \lambda^{\beta}$ where
$\beta=4/3$ and $\dot{m}_{E}$ is the accretion
rate in Eddington units \citep{cackett07}. We also expect the optical variations to be
smoother and have lower amplitude of variability than the UV
variations as they will come from a larger emission region. Both \cite{cackett07} and \cite{sergeev06} find lags consistent with $\beta=4/3$ between various optical bands. However neither study included X-ray data.

A number of observers have studied the relationship between the X-ray and optical wavebands, mostly combining ground based optical observations with space based X-ray observations from RXTE \cite[e.g.][]{uttley03_5548,suganuma06,
  arevalo08_2251, arevalo09, breedt09, breedt10, lira11, cameron12}. These observations have almost all
shown strong X-ray/optical correlations on short timescales (weeks - months) with the optical lagging the X-rays by $\sim$1~d. However, although the observations from the sample as a whole strongly support the conclusion that the optical lags the X-rays, in no individual case is the uncertainty on the lag small enough to be absolutely sure that the optical does lag.

Recent monitoring campaigns with Swift \cite[e.g.][]{shappee14, mch14, edelson15, troyer16, fausnaugh16, edelson17_4151} have greatly improved the measurement of lags between the X-ray, UV and optical bands and have therefore significantly improved our understanding of the origin of UV/optical variability in AGN. However these observations have also highlighted questions about the structures of accretion discs, of the importance of the BLR in producing reprocessed UV and optical emission and about whether X-rays from the central corona or maybe hard UV emission from the inner edge of the accretion disc are driving the longer wavelength UV/optical variability.
The above campaigns all show that the longer wavelength Swift UVOT bands lag behind the shortest wavelength Swift UV band (UVW2; 193 nm) in a manner which is in agreement with the short timescale (weeks/months) UV/optical variability of AGN being produced by reprocessing of radiation of shorter wavelength than UVW2 and coming from a compact region near the black hole. 
However, \cite{mch14} noted that if all of the reprocessing is being carried out by a surrounding accretion disc, that the disc is either hotter than or larger than we should expect, assuming the SS model and given the mass and accretion rate of the target AGN.
All subsequent papers \cite[e.g.][]{fausnaugh16} found a similar result. These observations were consistent with microlensing observations \cite[e.g.][]{morgan10,dai10,mosquera13} which had already pointed out a similar disc size discrepancy.
It was also clear  \cite[e.g.][]{edelson15, fausnaugh16} that the lag in the u-band was longer than that in surrounding bands, indicating that the BLR was also contributing to the lags.

\begin{figure*}
\includegraphics[width=180mm,height=240mm,angle=0]{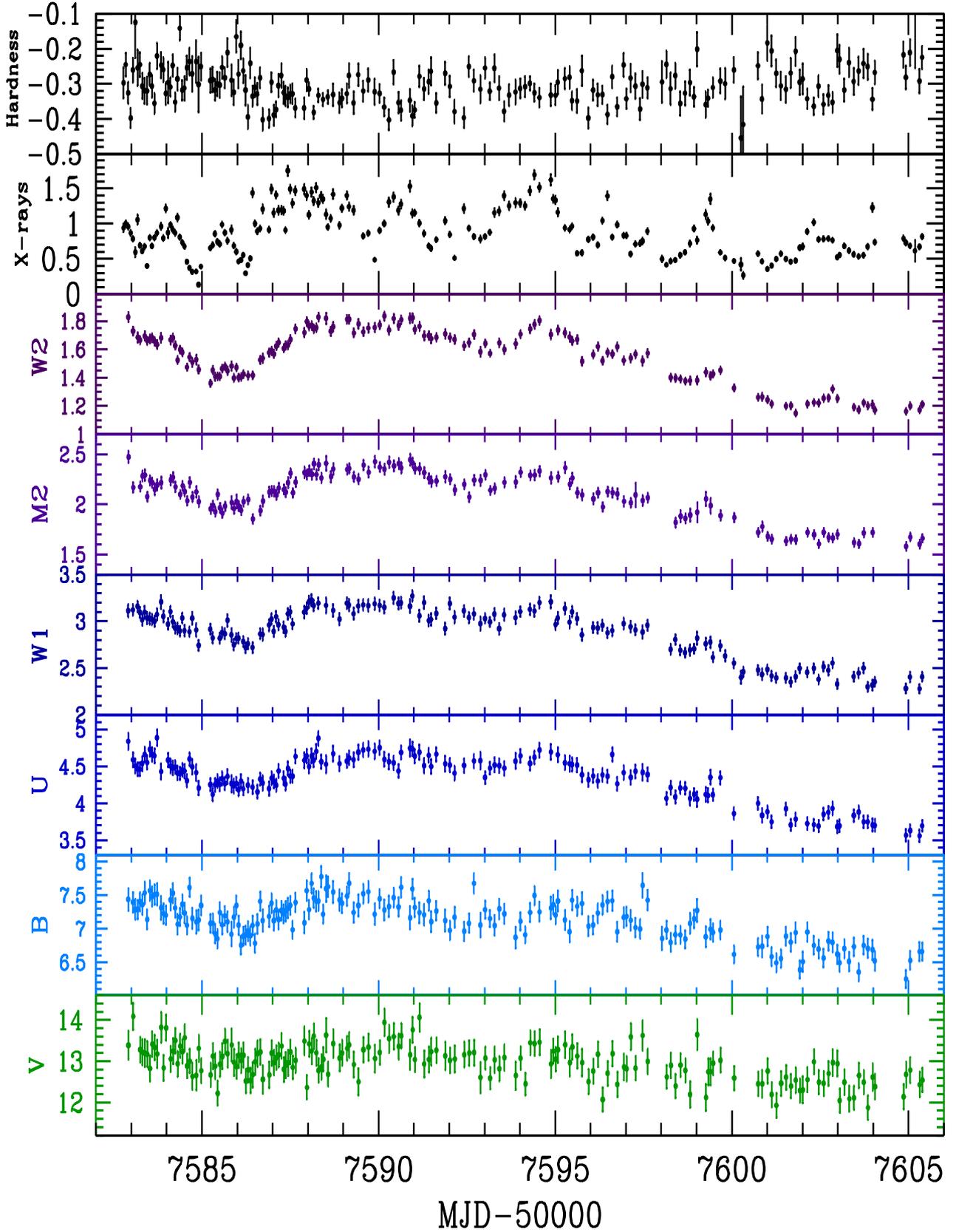}
\vspace*{-13mm}
\caption{XRT and UVOT lightcurves of NGC~4593. The top panel is the hardness ratio defined as H-S/H+S where H is the 2-10 keV count rate and S is the 0.5-2 keV count rate. The second from top panel is the 0.5-10 keV count rate for each visit. The lower panels are the UVW2 through to V-band fluxes in units of mJy.
}
\label{fig:lcs_8pan}
\end{figure*}

There has additionally been the concern \cite[e.g][]{berkley00,arevalo08_2251} that the optical lightcurves do not look as expected if they arise from reprocessing of X-ray emission from a small central corona, e.g. of size similar to that which we measure from microlensing observations, i.e. $\ltsim 10 R_{G}$ \citep{dai10,mosquera13}, or from X-ray low/high energy reverberation, ie $\sim4 R_{g}$ \citep{emmanoulopoulos14,cackett14}.  The observed optical lightcurves are smoother than expected and an insufficient fraction of the X-ray emission hits the disc to power the optical variability.  
Larger coronal sizes are required. \cite{gardner16} proposed an alternative model in which the X-ray emission does not directly impact on the outer disc but mainly heats up the very inner edge of the disc, which then inflates and re-radiates at hard UV wavelengths onto the outer disc. In this model there should be an additional lag between the X-ray and UVW2 emission, over and above that expected from an extrapolation of the longer wavelength lags down to the X-ray waveband. This additional lag would correspond to the thermal timescale for the incident X-ray heating to pass through the inner disc to the re-radiation surface. \cite{gardner16} note the existence of such a lag when the unfiltered X-ray and UVW2 observations of NGC~5548 are compared. However \cite{mch14} do not see any additional lag in NGC~5548 if those lightcurves are filtered to remove variations on timescales longer than 20d. In NGC~4151 \cite{edelson17_4151} see a very large excess lag between the X-ray and UVW2 bands. Unlike in NGC~5548, the excess lag in NGC~4151 is strongly energy dependent, with the highest energy X-rays having the largest lag. NGC~4151 is the most absorbed of the few AGN whose lags have been well studied so far and so the energy dependence may be a function of scattering in the absorbing medium. 

So far the number of AGN with accurately measured lags is small. With Swift, lags have been measured well in NGC~5548 \citep{mch14, edelson15, fausnaugh16} and NGC~4151 \citep{edelson17_4151} and less thoroughly in NGC~2617 \citep{shappee14} and NGC~6814 \citep{troyer16}.  With XMM-Newton lags have been measured in NGC~4395 between the X-ray and one UV (UVW1) and one optical (g) band \citep{mch16}.  The AGN in which lags have been measured well so far have been of relatively low accretion rate (NGC~4395 \me $\sim 0.0012$, NGC~4151 \me $\sim 0.021$ , NGC~5548 \me $\sim 0.048$). We therefore proposed for Swift monitoring of the somewhat higher accretion rate NGC~4593 (\me $\sim0.081$), allowing us to investigate the importance of accretion rate in determining disc and BLR structure.

In Section~\ref{sec:swiftobs} we present the Swift observations and lightcurves. The X-ray spectrum is relevant to understanding of lags and disc structure as, in the model of \cite{gardner16}, the tail of a luminous hard UV emission component may be expected to show up in at low X-ray energies. Thus in Section~\ref{sec:xspec} we discuss the time average Swift X-ray spectrum and X-ray spectral variability. In Section~\ref{sec:correlations} we discuss the relationships, and lags, between the X-ray and the various UV/optical bands. In Section~\ref{sec:disc} we compare the lags and the observed UVW2 lightcurve with the predictions from reprocessing of X-rays by a simple SS accretion disc. In Section~\ref{sec:memecho} we present a more sophisticated maximum entropy modelling of the X-ray/UV/optical lightcurves to derive reprocessing functions which are more complex than that expected from a simple accretion disc and which indicate the importance of reprocessing from the BLR. This topic is addressed in detail in a paper by \cite{cackett17} based on parallel Hubble Space Telescope (HST) observations of NGC~4593. In Section~\ref{sec:discussion} we compare the present lag measurements of NGC~4593 with those of other AGN and note broadly similar (scaled) lags between the UV and optical bands but differences in the X-ray/UV lags. We draw some brief conclusions regarding the inner structure of AGN.

Just before submission of this paper, the Swift data from this programme were
published by another group \citep{pal17}. Although there are some similarities, their analysis and
conclusions differ from ours in a number of respects, as we shall note below.

\section{SWIFT Observations}
\label{sec:swiftobs}

Swift observed NGC~4593 almost every orbit (96min) for 6.4 days from 13 to 18 July 2016 and thereafter every second orbit for a further 16.2d. Each observation totalled approximately 1~ks although observations were often split into two, or sometimes more, visits. 
The SWIFT X-ray observations are made by the X-ray Telescope \cite[XRT,][]{burrows05} and UV and optical observations are made by the UV and Optical Telescope \cite[UVOT,][]{roming05}.
In total 194 visits satifying standard good time criteria,
such as rejecting data when the source was located on known bad pixels, 
(e.g. see \url{https://swift.gsfc.nasa.gov/analysis/}
\url{xrt_swguide_v1_2.pdf}), 
were made. 
The XRT observations were carried out in photon-counting (PC) mode and the UVOT observations were carried out in image mode. X-ray lightcurves in a variety of energy bands were produced using our own Southampton pipeline which is based upon the standard Swift analysis tasks as described in \cite{cameron12}. We made flux measurements for each visit thus providing the best available time resolution. In addition, for comparison, a broad band 'snapshot' X-ray lightcurve (ie one flux point per visit) was produced using the Leicester Swift Analysis system \citep{evans07_lcs}, which was almost identical to our own snapshot lightcurve.  X-ray data are corrected for the effects of vignetting and aperture losses and data with large flux error ($>0.15$ count s$^{-1}$) are rejected.

During each X-ray observation, measurements were made in all 6 UVOT filters, using the 0x30ed mode which provides exposure ratios, for the UVW2, UVM2, UVW1, U, B, and V bands, of 4:3:2:1:1:1. 
UVOT lightcurves with the same time resolution were made using the Southampton system and also, independently, using a system developed by \cite{gelbord17}.
The latter system includes a detailed comparison of UVOT `drop out' regions, as first discussed in observations of NGC~5548 \citep{edelson15}. When the target source is located in such regions the UVW2 count rate is typically 10-15\% lower than in other parts of the detector. The drop in count rate is energy dependent, being greatest in the UVW2 band and least in the V-band. The new drop-out box regions are based on intensive Swift observations of three AGN, i.e. NGC~5548 \citep{edelson15}, NGC~4151 \citep{edelson17_4151} and the present observations of NGC~4593.  Observations falling in drop-out regions were rejected. We also searched for observations where the fluxes in the 6 UVOT bands showed a particularly red spectral slope. Such observations almost always fell within a drop-out region and were also rejected.

The resultant light curves are shown in Fig.~\ref{fig:lcs_8pan}. We see a close correspondence between all UVOT bands and a reasonable correspondence between the X-ray and UVOT bands. In Fig.~\ref{fig:lcs_8pan} we also show the X-ray hardness. The correspondences between these lightcurves are discussed in the following sections.

\section{X-ray Spectrum and Variability}
\label{sec:xspec}

\begin{figure}
\hspace*{-10mm}
\includegraphics[width=60mm,height=85mm,angle=270]{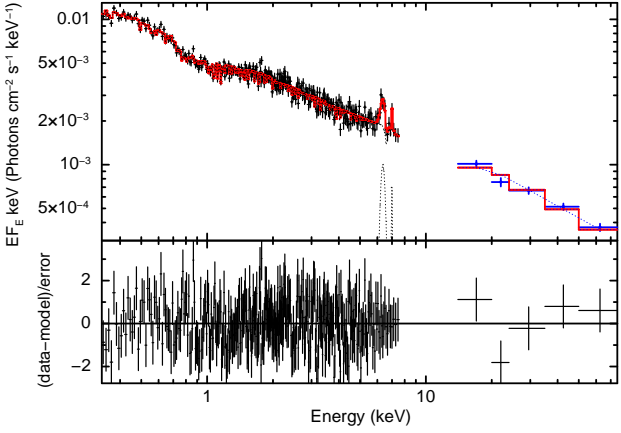}
\caption{Best fit, with residuals, to the time averaged XRT and BAT spectrum. The fit parameters are given in Table~\ref{tab:parm}.}
\label{fig:xspec}
\end{figure}

\subsection{Time Averaged X-ray Spectrum}
The X-ray spectrum can provide information which is relevant to understanding the inner geometries of AGN and so we have fitted models with {\sc xspec} to the Swift XRT time averaged spectrum derived from the observations shown in Fig.~\ref{fig:lcs_8pan}. The 2-8 keV spectrum is fitted well by a simple power law, with line of sight Galactic column cold hydrogen absorber of $1.89 \times 10^{20}$ cm$^{-2}$, together with a broad Gaussian line at 6.4 and a narrow line at $\sim7$ keV. The photon index of the power law is $\Gamma=1.68 \pm 0.02$. When this model is extended down to 0.3 keV a large negative residual is seen centred on $\sim1$ keV which is consistent with the presence of warm absorbers as detected by previous observers using data from XMM-Newton \citep{brenneman07_4593} and combined data from XMM-Newton and NuStar \citep{ursini16_4593_MN}. Following these studies we add two warm absorbers which results in a good fit to the 0.3-8 keV spectrum with $\Gamma=1.74 \pm 0.01$. This fit is similar to that of Brenneman et al ($\Gamma=1.75^{+0.02}_{-0.03}$) except that, unlike them, we do not require an additional 'soft-excess' component at low energies. 

NGC~4593 is detected in the Swift BAT 70 month survey \citep{baumgartner13_bat_survey}. The spectrum is described as a power law with $\Gamma=1.84^{+0.07}_{-0.08}$ and flux (14-195 keV) of $8.8 \pm 0.05 \times 10^{-11}$ ergs cm$^{-2}$ s$^{-1}$
(\url{https://swift.gsfc.nasa.gov/results/bs70mon/}
\url{SWIFT_J1239.6-0519}). During the monitoring reported here the average 0.3-10 keV flux was $\sim 4.5 \times 10^{-11}$ ergs cm$^{-2}$ s$^{-1}$. To extrapolate a power law to the BAT energy band, and obtain the observed BAT flux requires $\Gamma=1.73$, indicating that the long term average X-ray luminosity of NGC~4593 has not changed noticeably. Including the BAT data into our spectral fit without any change of normalisation, and assuming a simple power law without any high energy cut-off, steepens the overall fit slightly ($\Gamma=1.82^{+0.04}_{-0.05}$). The overall best fit is shown in Fig.~\ref{fig:xspec} and the fit parameters are given in Table~\ref{tab:parm}.

\begin{table}
 \centering
 \caption{X-Ray Spectral Fit Parameters}
 \label{tab:parm}
\begin{tabular}{cc}
\hline 
Spectral & Fitted    \\
Parameters & values  \\
\hline
N$_{H,\rm tbabs}$ [10$^{22}$ cm$^{-2}$] & 0.0189 (f) \\ 
$NH_{\rm WA1,\rm zxipcf}$ [10$^{22}$ cm$^{-2}$] & 1.22$^{+0.12}_{-0.13}$ \\
$\log \zeta_{\rm WA1,\rm zxipcf}$ & 0.87$^{+0.11}_{-0.12}$ \\
f$_{\rm WA1,\rm zxipcf}$ & 0.47$^{+0.02}_{-0.03}$ \\
NH$_{\rm WA2,\rm zxipcf}$ [10$^{22}$ cm$^{-2}$] & 33.9$^{+7.8}_{-12.3}$ \\
$\log \zeta_{\rm WA2,\rm zxipcf}$ & 3.51$^{+0.03}_{-0.02}$ \\
f$_{\rm WA2,\rm zxipcf}$ & 1 (f) \\
$\Gamma_{\rm zpowerlw}$ & 1.82$^{+0.04}_{-0.05}$ \\
F$_{\rm zpowerlw}$ [10$^{-11}$ ergs s$^{-1}$ cm$^{-2}$] & 4.73$^{+0.05}_{-0.08}$ \\
E$_{\rm K\alpha,\rm zgauss}$ [keV] & 6.42$^{+0.03}_{-0.04}$ \\
$\sigma_{\rm K\alpha,\rm zgauss}$ [keV] & 0.21$^{+0.04}_{-0.06}$ \\
F$_{\rm K\alpha,\rm zgauss}$ [10$^{-11}$ ergs s$^{-1}$ cm$^{-2}$] & 0.072$^{+0.013}_{-0.024}$ \\
Q$_{\rm K\alpha,\rm zgauss}$ [eV] & 373$^{+63}_{-57}$ \\
E$_{\rm K\beta,\rm zgauss}$ [keV] & 7.06$^{+0.07}_{-0.08}$ \\
$\sigma_{\rm K\beta,\rm zgauss}$ [keV] & 0.05 (f) \\
F$_{\rm K\beta,\rm zgauss}$ [10$^{-11}$ ergs s$^{-1}$ cm$^{-2}$] & 0.023$^{+0.014}_{-0.008}$ \\
Q$_{\rm K\beta,\rm zgauss}$ [eV] & 86$^{+40}_{-39}$ \\
F$_{0.3-2}$ [10$^{-11}$ ergs s$^{-1}$ cm$^{-2}$] & 1.39$^{+0.05}_{-0.04}$ \\
F$_{2-10}$ [10$^{-11}$ ergs s$^{-1}$ cm$^{-2}$] & 3.43$^{+0.04}_{-0.03}$ \\
F$_{10-100}$ [10$^{-11}$ ergs s$^{-1}$ cm$^{-2}$] & 5.52$^{+0.06}_{-0.11}$ \\
$\chi^2$/dof & 374/342 \\
\hline
\end{tabular}
\vspace*{3mm}

\begin{minipage}{80mm}
Best fit model parameters from the simultaneous fitting of \swift{}/XRT and \swift{}/BAT energy spectra using the model \textsc{const $\times$ tbabs $\times$ [zxipcf*zxipcf*zpowerlw+zgauss+zgauss]}. N$_{H,\rm tbabs}$ is the Galactic absorption colum density, $NH_{\rm WA1,\rm zxipcf}$, $\log \zeta_{\rm WA1,\rm zxipcf}$ and f$_{\rm WA1,\rm zxipcf}$ are the column density, ionization parameter and partial covering fraction respectively due to the first warm absorber component while the same parameters for the second warm absorber component are denoted by `WA2'. $\Gamma_{\rm zpowerlw}$ is the photon powerlaw index. E$_{\rm K\alpha,\rm zgauss}$, $\sigma_{\rm K\alpha,\rm zgauss}$ and Q$_{\rm K\alpha,\rm zgauss}$ denote the line energy, line width and the equivalent width of the Fe K$\alpha$ emission line while E$_{\rm K\alpha,\rm zgauss}$, $\sigma_{\rm K\alpha,\rm zgauss}$ and Q$_{\rm K\alpha,\rm zgauss}$ denote the line energy, line width and the equivalent width of the Fe K$\beta$ emission line respectively. F$_{\rm zpowerlw}$, F$_{\rm K\alpha,\rm zgauss}$ and F$_{\rm K\beta,\rm zgauss}$F$_{\rm cutoffpl}$ are the fluxes due to $\textsc{zpowerlw}$ and two $\textsc{zgauss}$ models respectively in the energy range 0.3-10.0 keV. F$_{0.3-2.0}$, F$_{2.0-10.0}$, F$_{10.0-100.0}$ are unabsorbed fluxes in the energy range 0.3-2 keV, 2-10 keV, and 10-100 keV respectively.
\end{minipage}
\end{table}

To the best-fit model we added, separately, a bremmstrahlung ('zbremss') and a Comptonisation ('comptt')  component as used by \cite{brenneman07_4593} to describe a soft excess which they find in their XMM-Newton data. In our Swift data the normalisations of both components are consistent with zero and the $1 \sigma$ upper limit on the 0.3-10 keV fluxes for these two components are $6 \times 10^{-16}$ and $3.4 \times 10^{-17}$ ergs cm$^{-2}$ s$^{-1}$ respectively (compared to $2 \times 10^{-14}$ and $6.46 \times 10^{-14}$ ergs cm$^{-2}$ s$^{-1}$ respectively from Brenneman et al). 
We note that Brenneman et al find variation in the soft excess between different observations so it is possible that we observed with Swift when the soft excess was particularly faint.
\cite{pal17} present a 0.3-7 keV Swift XRT spectrum. It is of much lower S/N than that presented here,
possibly being only from a single 1~ks observation. They are therefore able only to fit
to a power law, whose slope is not well constrained, and a black body. They do not include the warm absorbers or the iron lines.

\begin{figure}
\hspace*{-10mm}
\includegraphics[width=60mm,height=85mm,angle=270]{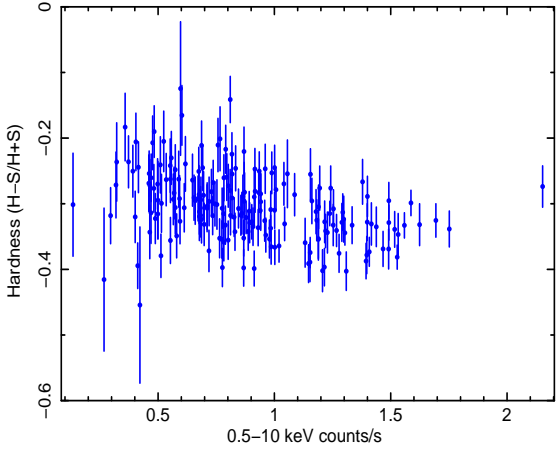}
\caption{The X-ray Hardness as a function of the 0.5-10 keV count rate. The hardness is defined as (H-S)/(H+S), where S=0.5-2 keV and H=2-10 keV count rate.}
\label{fig:hardnessvscounts}
\end{figure}

\begin{figure}
\hspace*{-10mm}
\includegraphics[width=60mm,height=85mm,angle=270]{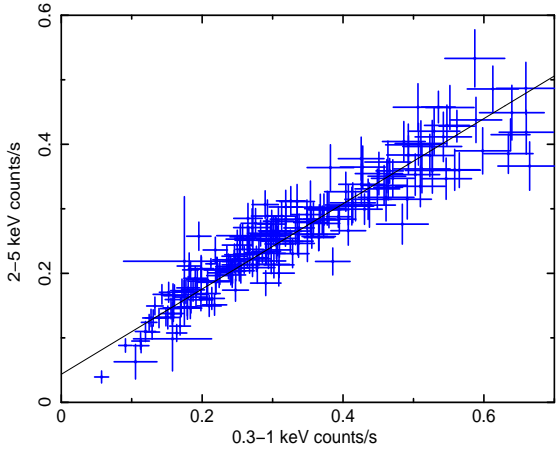}
\caption{The 0.3-1 keV count rate plotted against the 2-5 keV count rate}
\label{fig:xrayfluxflux}
\end{figure}

\subsection {X-ray Spectral Variability and X-ray Energy Dependence of Lags}

In Swift observations of NGC~4151, \cite{edelson17_4151} found large differences in the lag measured between different X-ray bands and the UVW2 band. Although \cite{mch14} did not find any significant differences between the 0.5-2 keV vs UVW2 lags and the 2-10 keV vs UVW2 lags in NGC~5548, the possibility exists that different X-ray bands may come from different locations and so give rise to different lags. 

We have therefore made lightcurves in a variety of narrow and broad X-ray energy bands and have searched for lags between them using a variety of techniques. We can find no measureable lags. For example, using {\sc JAVELIN} \citep{zu11_javelin,zu13_javelin}, we find that the 0.5-2 keV band lags the 2-10 keV band by $-0.001^{+0.003}_{-0.004}$d. Similarly the lag of the 2-10 keV band by the 0.3-1 keV band is $-0.002^{+0.004}_{0.005}$d. 

As an additional method of searching for differences between X-ray bands we have calculated the hardness ratio. The hardness ratio is defined as (H-S)/(H+S), where here the hard (H) band is 2-10 keV and the soft (S) band is 0.5-2 keV. We plot this ratio as a function of time in the top panel of Fig.~\ref{fig:lcs_8pan}. There is little variation. In Fig.~\ref{fig:hardnessvscounts} we plot the hardness ratio against broad band (0.5-10 keV) count rate. Above 0.5 counts/s there is a very slight softening of the spectrum with increasing count rate. This spectral softening is similar to that found in NGC~4593 by \cite{ursini16_4593_AN} and for AGN in general by \cite{sobolewska09}.
Although they use 0.3-1.5 keV and 1.5-10 keV as their soft and hard band respectively, and define hardness ratio as H/S rather than the (H-S)/(H+S) used here, similar results are shown by \cite{pal17}.

We note here that below 0.5 counts s$^{-1}$ (0.5-10 keV) data are limited so it is not clear whether the suggestion of softening with decreasing count rate, at the lowest count rates, is real or not. A softening with decreasing count rate at the lowest count rates has been seen in NGC~1365 \citep{connolly14} and attributed to unabsorbed (ie steep spectrum)
X-rays scattered from an accretion disc wind which are still visible even when the direct X-ray emission is heavily absorbed.

To determine whether the slight softening with increasing luminosity represents a real change in the underlying spectrum we plot, in Fig.~\ref{fig:xrayfluxflux}, the 0.3-1 keV count rate against the 2-5 keV count rate.  Similar flux-flux plots have been used elsewhere to investigate the reasons behind flux related X-ray spectral variations \cite[e.g.][]{taylor03}. Above a 0.3-1 keV count rate of $\sim 0.15$ counts s$^{-1}$, there is a strong linear relationship between the count rates in the two bands with an  extrapolation to zero 0.3-1 keV count rate giving a small residual count rate in the harder band.
The data are again insufficient to determine whether there is any real deviation from this relationship at the very lowest count rates. 

Five combined XMM-Newton and NuStar observations have been fitted with a multi-component model including a power law and a high energy cut-off. From this modelling it is stated that the photon index varies by $\sim0.25$ over a factor 3 in luminosity \citep{ursini16_4593_MN}. The NuStar observations extend to a higher energy than the Swift XRT observations but, within the XRT observations, the strong linear relationship between the hard and soft X-ray count rates shows that there is no change of spectral shape of the varying component as a function of luminosity over the large majority of the flux range observed. 
The weak softening of the overall spectrum with increasing luminosity shown in Fig.~\ref{fig:hardnessvscounts} is most simply explained as the combination of a small constant component of hard spectrum together with a varying soft spectrum component. The hard component may be a reflection component from the disc or BLR.  
We therefore conclude that, unlike in NGC~4151, all of the XRT energy band varies simultaneously and so, to increase S/N, we hereafter use the 0.5-10 keV band unless stated otherwise.

\section{X-ray / UV-Optical Correlations}
\label{sec:correlations}

\subsection{X-ray/UVW2 Discrete Correlation Function}

\begin{figure}
\hspace*{-10mm}
\includegraphics[width=60mm,height=85mm,angle=270]{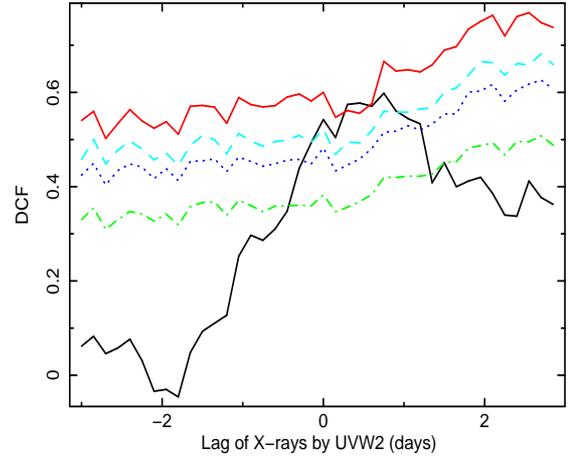}
\caption{ Solid black line - Discrete cross correlation function between the X-ray
  and UVW2 lightcurves shown in
  Fig.~\protect\ref{fig:lcs_8pan}. The 68\% (dot-dash, green), 90\% (dotted, blue) 95\% (dashed turquiose) and
  99\% (solid red) confidence levels are also shown. }
\label{fig:xw2_dcf}
\end{figure}

Visual inspection of the X-ray and UVW2 lightcurves indicates that most of the X-ray flux variations on $\sim$day timescales have counterparts, though of lesser fractional variability (Table~\ref{tab:lags}), in the UVW2 lightcurve. 
As a basic method of quantifying the relationship between these two bands we show,
in Fig~\ref{fig:xw2_dcf}, the discrete correlation function \cite[DCF,][]{edelson88} between these two bands together with simulation-based confidence contours. We see that a correlation exists between these two bands at greater than 99\% confidence. This degree of confidence between the observed X-ray and UVW2 lightcurves, without any filtering to remove long term trends which often distort DCFs, is higher than in all other previous intensive Swift AGN monitoring programs \cite[e.g.][]{mch14,edelson15,edelson17_4151}.
The peak lag corresponds to the UVW2 lagging the X-rays by about half a day. The exact value of the lag will be considered in more detail later.

\begin{figure*}
\hspace*{-5mm}
\includegraphics[width=52mm,height=60mm,angle=0]{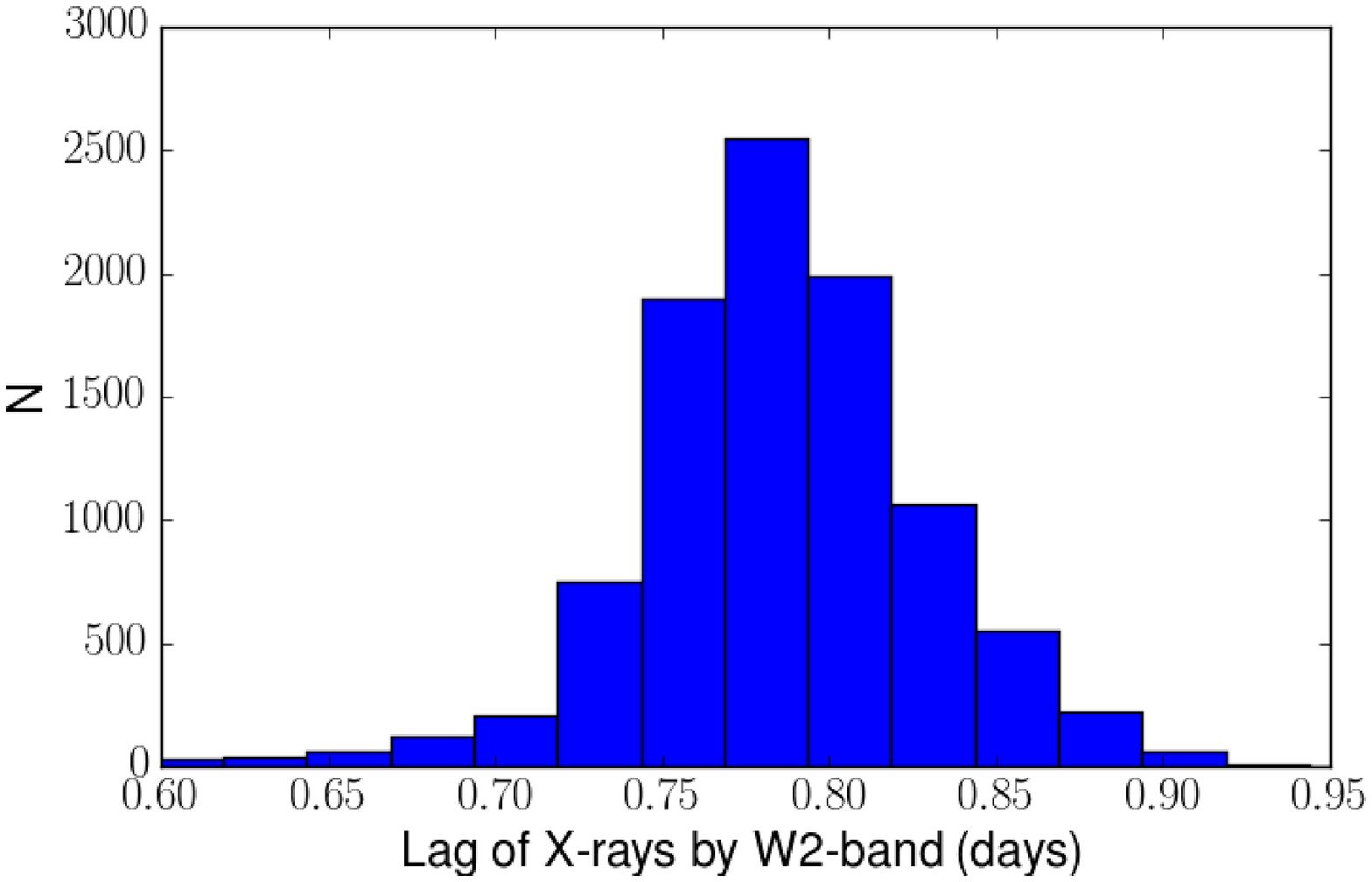}
\includegraphics[width=52mm,height=60mm,angle=0]{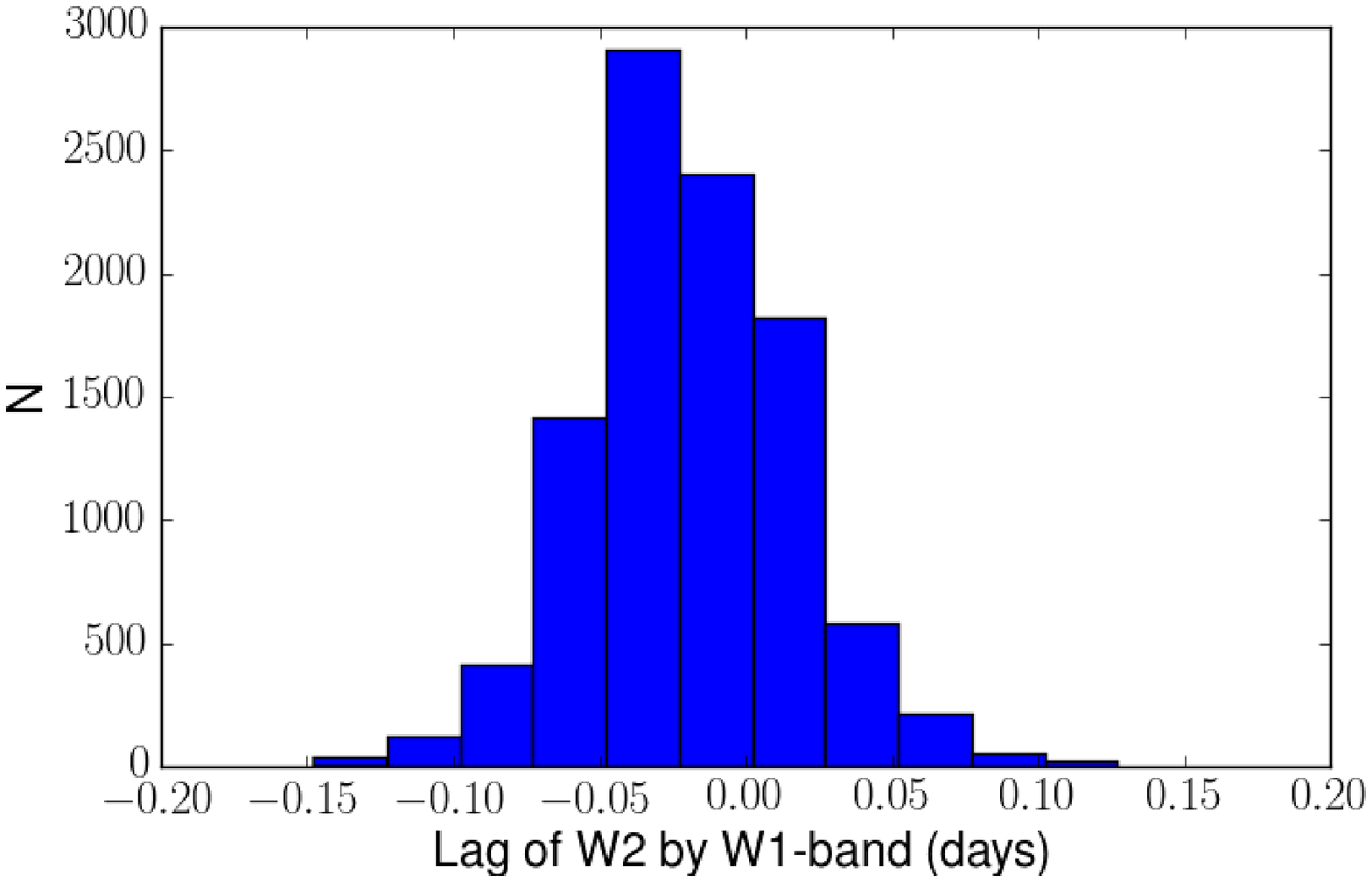}
\includegraphics[width=52mm,height=60mm,angle=0]{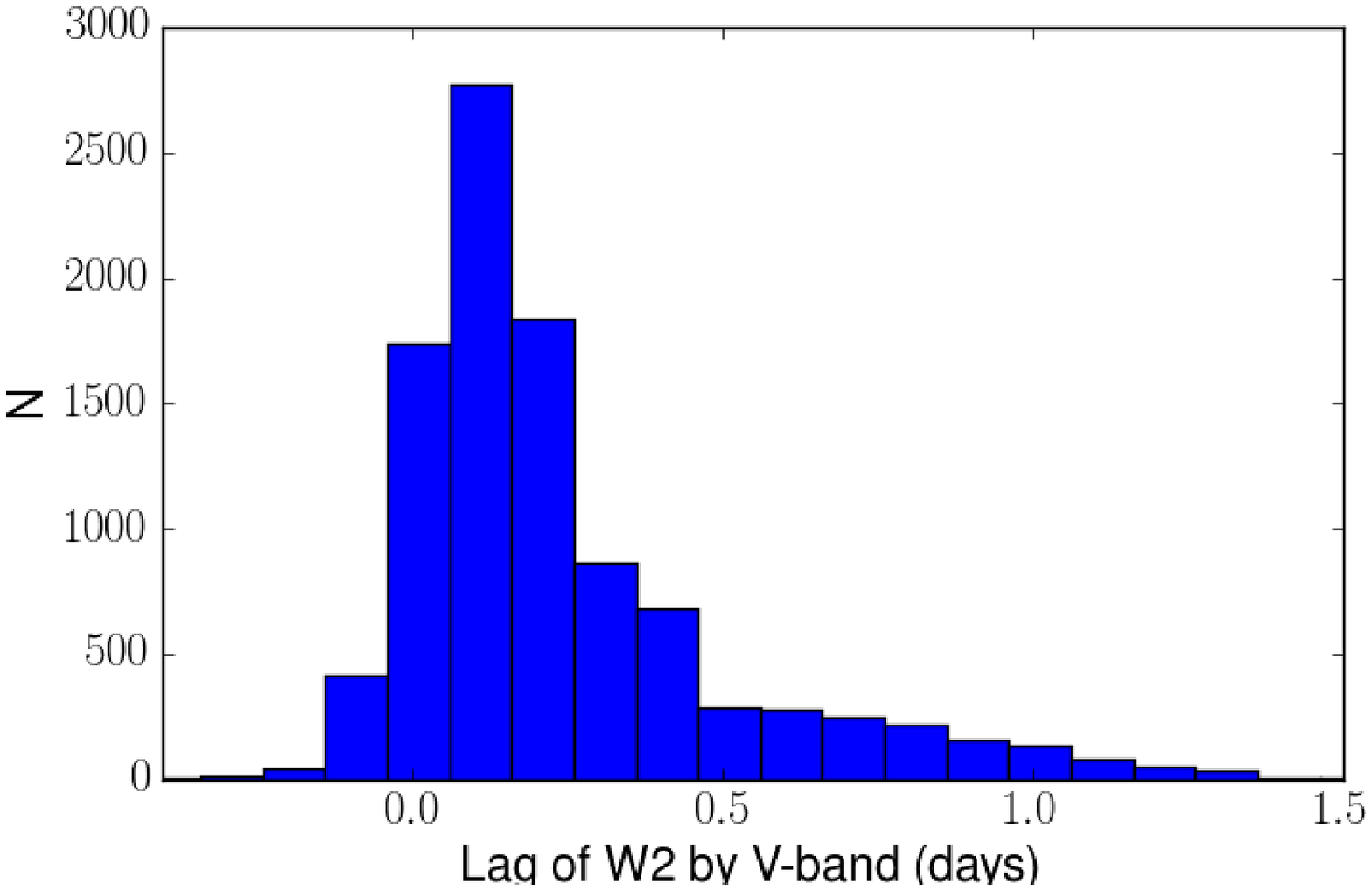}
\caption{Lag probability distributions from {\sc JAVELIN} for (left panel) UVW2 relative to the X-rays, (centre) UVW1 relative to UVW2 and (right) V-band relative to UVW2.}
\label{fig:javlagdistributions}
\end{figure*}

The confidence contours are calculated in broadly the same way as in our previous papers on X-ray/optical correlations \cite[e.g.][]{breedt09}. X-ray lightcurves are simulated with the same variability properties, eg power spectral density (PSD) and count rate probability density function, as for the observed X-ray lightcurve. The N\% confidence levels are defined such that if correlations are performed between the observed UVW2 data and randomly simulated X-ray lightcurves, only (100-N)\% of the correlations would exceed those levels. The confidence levels are appropriate to a single trial, ie a search at zero lag, approximately what we are searching for here. For a search over a wide lag range where the expected lag was unknown, the confidence levels would be reduced by an amount depending on the ratio of the lag range being searched to the width of the expected correlation function (ie $(ACF\_Width_{X}^{2} + ACF\_Width_{O}^{2})^{1/2}$ where $ACF\_Width_{X}$ and $ACF\_Width_{O}$ are the half-widths of the X-ray and optical autocorrelation functions. If this ratio was $R$, then the single-trial confidence level of (100-N)\% would reduce to (100-RN)\%.  We note that the previous lightcurve simulation method depended only on the parameters of the X-ray PSD, using the method of \cite{timmer95}. The present lightcurve simulation method follows \cite{emmanoulopoulos13b}, with code available from \cite{connolly15}, which also takes account of the count rate probability density function (PDF). Unlike the method of \cite{timmer95} which can only produce Gaussianly distributed lightcurves, this method can produce lightcurves with any PDF, including the highly non-linear lightcurves seen in the Gamma-ray and TeV bands.

\begin{figure}
\hspace*{-10mm}
\includegraphics[width=50mm,height=85mm,angle=270]{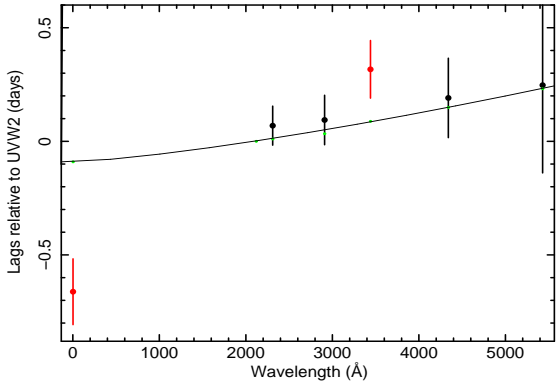}
\caption{ Lags relative to UVW2 derived from the centroids of the distribution using the FR/RSS method \protect\cite{peterson98}. The thin line is a power law fit with index 4/3 to the model lags (shown as small green dots). Here the model lag is the time for half of the light to be received. 
}
\label{fig:centroidlags}
\end{figure}

\begin{figure}
\hspace*{-10mm}
\includegraphics[width=50mm,height=85mm,angle=270]{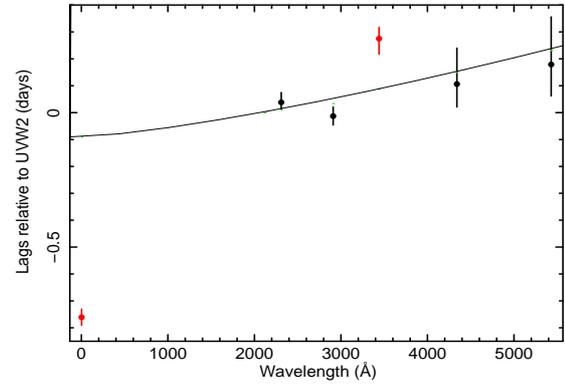}
\caption{Lags relative to UVW2 derived using {\sc JAVELIN} \protect\cite{zu11_javelin,zu13_javelin}. The thin line is a power law fit with index 4/3 to the model lags (shown as small green dots). Here the model lag is the time for half of the light to be received.}
\label{fig:javlags}
\end{figure}

The level and detailed shape of the confidence contours do change, though only very slightly (see similar conclusion in \cite{edelson17_4151}), depending on the model chosen for the X-ray PSD. Here we used the standard bending power law model of \cite{mch04} with Poisson noise and fixed the low frequency PSD slope at -1. The PSD derived here from the Swift XRT observations is still not very well constrained but a high frequency slope of -1.8 was measured, similar to that (-2.2) found by \cite{summons07_phd} from RXTE and XMM-Newton observations. However even quite large changes in PSD model do not change the confidence contours greatly. In almost all tests the peak of the DCF reached a significance level of between $\sim$95 and 99 per cent.

DCFs for the relationship between the UVW2 band and the other UVOT bands all show a strong peak near zero lag with very high significance ($>99.9$ per cent confidence) and are not shown here. The DCF does not provide a particularly accurate measurement of the value of the lag and so the lag measurements are derived using other methods, below.

\subsection{X-ray/UV/Optical Interband lags}

All lags are measured relative to the UVW2 band as the UVW2 band provides the highest significance detections of any of the UVOT bands. Lags, with errors, were measured both using {\sc JAVELIN} \citep{zu11_javelin,zu13_javelin}, as previously demonstrated by \cite{shappee14,pancoast14,mch14}, and using the `Flux Randomization/Random Subset Selection' (FR/RSS) method with the interpolation cross-correlation function, \cite[ICCF,][]{peterson98}. The median lag values produced by {\sc JAVELIN} are usually close to the value produced by the FR/RSS method but, as noted by \cite{fausnaugh16}, the uncertainties are usually smaller. 

\begin{figure}
\includegraphics[width=85mm,height=60mm,angle=0]{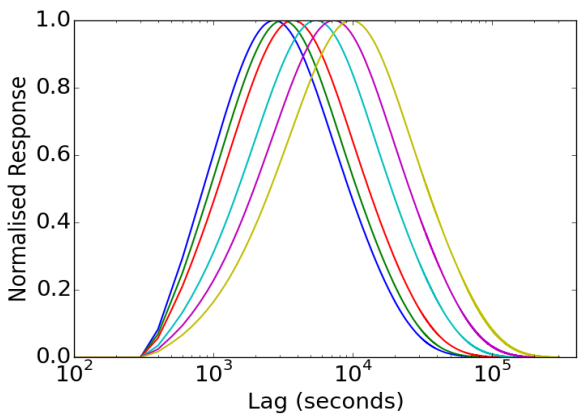}
\caption{Impulse response functions, normalised to the peaks, for an accretion disc surrounding a Schwarzschild black hole of the mass of NGC~4593. It is assumed that the disc reaches to the last innermost stable circular orbit. The X-ray source is assumed to be a point source located 6~$R_{G}$ above the axis of the black hole. The responses increase in lag through the UVOT bands from UVW2 to V-band.}
\label{fig:responses}
\end{figure}

The differences in the uncertainties arise from the different methodologies employed. {\sc JAVELIN} directly compares many predicted ouput, or 'driven', lightcurves based on one input 'driving' lightcurve. The FR/RSS method produces a distribution of ICCFs derived from differing sampling of the observed driven and driving lightcurves. {\sc JAVELIN} uses a damped random walk (DRW) to interpolate whereas FR/RSS uses linear interpolation. The DRW has a PSD slope of -2 at high frequencies, similar to that observed in the X-ray band (Section \ref{sec:correlations}), and a slope of zero at low frequencies. The observed low frequency PSD slope is -0.8, with a PSD bend timescale of 3d \citep{summons07_phd}. Thus on the timescales of the lags measured here, the DRW is a reasonable description of the X-ray variability. However the best measured optical AGN PSD slopes, as measured from Kepler observations \citep{mushotzky11}, lie in the range -2.6 to -3.3. Although these values are steeper than the -2 of a simple DRW, the output 'driven' lightcurve in {\sc JAVELIN} is filtered by convolution with a top hat function and so, on timescales of interest to the lags, the PSD of the driven lightcurve should be steeper than -2. When comparing one UV/optical band against another, however, the DRW will not be a perfect description of the driving lightcurve. Also, the top-hat smoothing function used by {\sc JAVELIN} is not an exact version of the response functions show in Fig.~\ref{fig:responses}. 
The uncertainties shown here for both methods are statistical and do not include any systematic uncertainties arising from the deviation of the physical reality from the assumptions of the methods. A full discussion of the relative merits of the two methods is beyond the scope of this paper and so here we present results from both methods.

The FR/RSS method produces two alternative lag measurements, based either on the distribution of the values of the centroids of the individual ICCFs (usually measured at 80 per cent of the peak value), or on the distribution of the peak lag values. With asymmetric ICCFs, these distribution will differ and provide us with different information. Where an ICCF arises from the sum of a number of contributing
lags, the centroid provides us with an estimate of the average lag. The peak highlights the dominant contributor to the lags.

In Fig.~\ref{fig:centroidlags} we show the lags measured, from the centroids of the FR/RSS lag distributions. We do not show the lags derived from the peaks of the FR/RSS distributions but we give the values in Table~\ref{tab:lags}.
In Fig.\ref{fig:javlags} we show the lags measured using {\sc JAVELIN} which are more like the FR/RSS peak than centroid lags.
The lag distributions between the UVW2 and short wavelength bands (X-ray, UVM2) are symetrical, but tails to longer lags appear in the b and v bands (Fig.~\ref{fig:javlagdistributions}) indicating a possible secondary source of lags. 

\begin{table*}
\caption{F$_{var}$ and Lags (days) relative to UVW2 from various correlation methods.}
\begin{tabular}{lcccc}
  Band  & F$_{var}$ & {\sc JAVELIN} & FR/RSS Peak & FR/RSS Centroid \\
&&&&\\
0.5-10 keV & 0.376 & $-0.761^{+0.031}_{-0.030}$  & $-0.412 \pm 0.186$ & $-0.662 \pm 0.145$ \\
UVW2       & 0.125 & -  & - & - \\
UVM2       & 0.110 & $+0.038^{+0.037}_{-0.025}$ & $+0.057 \pm 0.083$ & $+0.069 \pm 0.086$ \\
UVW1       & 0.092 & $-0.013^{+0.034}_{-0.033}$ & $-0.013 \pm 0.078$ & $+0.094 \pm 0.109$ \\
U          & 0.070 & $+0.275^{+0.042}_{-0.058}$ & $+0.186 \pm 0.214$ & $+0.317 \pm 0.127$ \\
B          & 0.038 & $+0.106^{0.134}_{-0.085}$ & $+0.101 \pm 0.208$ & $+0.191 \pm 0.175$ \\
V          & 0.024 & $+0.179^{0.177}_{-0.117}$ & $+0.220 \pm 0.502$ & $+0.247 \pm 0.386$ \\
\end{tabular}
\label{tab:lags}
\end{table*}

Relative to UVW2, the other UV and optical band lags are very small. 
With all measurement methods we see that the u-band lag significantly exceeds any interpolation between the lags in bands on either side. As in NGC~5548 \citep{edelson15,fausnaugh16}, this excess lag is probably due to a contribution from the broad line region and is discussed in detail by \cite{cackett17}. The lag measured by {\sc JAVELIN} to the UVW1 band is actually negative, though is also consistent with a small positive value. Possibly there is some contribution from [CIII] from the BLR to the emission in UVW2 but not in UVW1.

\cite{pal17}, although finding lags which increase with wavelength, do not obtain the same lag values as those given here. The main difference between our analyses is that 
\citeauthor{pal17} use 1.5-10 keV as their reference band against which to measure lags in all other bands, from soft X-ray to V-band, whereas we measure all other UVOT band lags relative to UVW2. As we see in Fig.~\ref{fig:lcs_8pan}, and as seen in all previous Swift AGN monitoring campaigns, the UVW2 lightcurve has the highest S/N of any of the UVOT bands. Also the main change in the character of the variability is between the X-rays and UVW2 whereas the other UVOT bands are quite similar in character to UVW2. Thus the peak correlation strength between the X-rays and UVW2 is $\sim0.6$ (as shown also by \citeauthor{pal17}) whereas the peak correlation strength between UVW2 and the other UVOT bands is greater than 0.9. Thus lag measurements between UVW2 and the other UVOT bands have much smaller errors than lag measurements between the hard X-ray band and the UVOT bands. We also find no evidence for lags between hard and soft X-ray bands and so we measure a lag between UVW2 and the broad 0.5-10 keV X-ray band, which has higher S/N than just the 1.5-10 keV band. Thus we believe that the lags presented here are a more accurate measurement of the true lags than those presented by \citeauthor{pal17}.

In Figs~\ref{fig:centroidlags} and ~\ref{fig:javlags} we plot model lags on the assumption that the variable UV/optical emission is produced by reprocessing of X-ray emission on an accretion disc as described by \cite{shakura73}. As previously \citep[e.g.][]{mch14}, we take, as the model lag, the time for half of the reprocessed light to be received. We refer to this lag as the median lag.
The response functions for the various UVOT bands from which the lags are derived are shown in Fig.~\ref{fig:responses}. There are a number of assumptions in the derivation of the response functions and in the resultant lag estimation which are discussed below (Section~\ref{sec:disc}). Here we simply note that, if the model lags are measured relative to the UVW2 band, then the other observed UV and optical lags agree reasonably with the model. This conclusion differs slightly from our conclusion regarding the lags in NGC~5548 \citep{mch14} where, for the lag of the UVW2 band by the V-band, we stated that the observed lags were a factor of 2.25 larger than the model. There is a slight caveat that the fortran code which produced the model lags in \cite{mch14} is no longer operational so an entirely new python code was written which integrates the contributions from the different sections of the disc in a different way. The new code produces response functions which are very similar to the original code but the median lag time is $\sim15-20$ per cent longer. Given the complete independence of the two codes it is encouraging that they both produce very similar results. However, with the present code, we would have said that the observed UVW2-V band lag in NGC~5548 given in \cite{mch14} was $\sim1.9 \times$ longer than the model. We compare the UVW2-V band lags between different AGN in Section~\ref{sec:discussion}.

For either code, when we extrapolate the model back to the X-ray waveband, the observed UVW2 emission lags the X-rays by much more (factor $\sim6$) than expected from the model. Thus although the lags within the UV and optical bands are quite consistent with reprocessing of far-UV emission by an accretion disc, as proposed by \cite{gardner16}, 
a simple lamp-post point X-ray source directly illuminating only a surrounding accretion disc cannot explain the complete spectrum of multiband lags from X-ray to V-band. Possible explanations are discussed below (Sections~\ref{sec:disc} and ~\ref{sec:memecho}).

\subsection{UV-optical Spectral Variability}

The lightcurves shown in Fig.~\ref{fig:lcs_8pan}, and the fractional variances listed in Table~\ref{tab:lags}, show greater variability at shorter UVOT wavelengths. Is this difference a reflection of complex flux-related spectral variability or simply contamination by a galaxy component? We can investigate the origin of the variability using flux-flux analysis, similar to that employed above to investigate X-ray spectral variability. 

\begin{figure}
\hspace*{-5mm}
\includegraphics[width=80mm,height=85mm,angle=270]{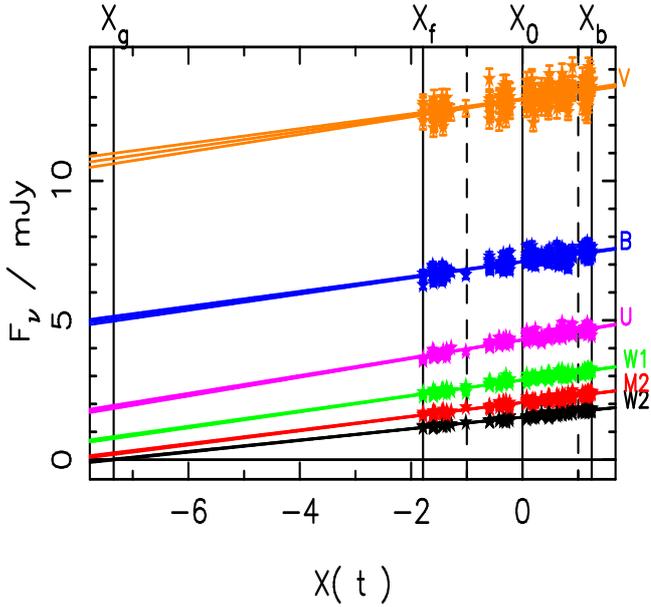}
\caption{$F_{\nu}(\lambda, t)$ vs. $X(t)$, as defined in equation~\ref{eqn:fv}, for each of the UVOT bands.}
\label{fig:uvotxrayfluxes}
\end{figure}

Here we fit the fluxes (here using $F_{\nu}$ in mJy) as a function of time, in each UVOT band, ie $F_{\nu}(\lambda, t)$, as 
\begin{equation}
F_{\nu}(\lambda, t) = A_{\nu}(\lambda) + R_{\nu}(\lambda) X(t) 
\label{eqn:fv}
\end{equation}
where $A_{\nu}(\lambda)$ is a constant component, representing the mean spectrum, $R_{\nu}(\lambda)$ is the rms spectrum and $X(t)$ is a dimensionless lightcurve such that $<X>=0$ and $<X^{2}>=1$. In Fig.~\ref{fig:uvotxrayfluxes} we plot $F_{\nu}(\lambda,t)$ against $X(t)$ for each of the UVOT bands. A clear linear response is seen in all cases, together with different constant offsets. This linear response shows, as in the earlier X-ray flux-flux analysis, that each uvot band is well described by a combination of a variable component whose spectrum does not change with luminosity, and a constant component. The 'bluer when brighter' variations that are apparent in the raw lightcurves are thus purely a result of dilution of a bluer variable component (from the accretion disc) and a redder component (from the host galaxy). \cite{pal17} present plots of the raw UVOT count rates against the 1.5-10 keV count rate, showing approximate correlations with a good deal of scatter. They do not attempt any spectral modelling.

\begin{figure}
\includegraphics[width=75mm,height=80mm,angle=270]{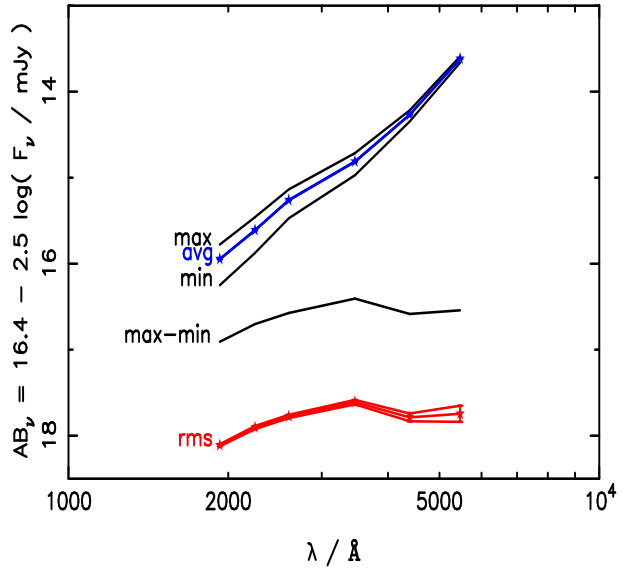}
\caption{The lower curve, labelled 'rms', is the rms spectrum of the variable uvot component, $R_{\nu}(\lambda)$, as defined in equation~\ref{eqn:fv} and derived from the slopes of the plots shown in Fig.~\ref{fig:uvotxrayfluxes}. The uncertainties are derived from the uncertainties on the slopes.
The middle curve, labelled 'max-min', represents the difference between the maximum and minimum observed UVOT fluxes. The upper curve, labelled 'avg' is the lower limit on the constant host galaxy component of the UVOT fluxes, derived from the intercepts of the curves shown in Fig.~\ref{fig:uvotxrayfluxes} at the point where the UVW2 flux is zero.
}
\label{fig:uvotspectrum}
\end{figure}

The slopes of $F_{\nu}(\lambda, t)$ vs $X(t)$ give $R_{\nu}$, the spectrum of the variable component. This rms spectrum, with upper and lower limits derived from the slope uncertainties, is show in Fig.~\ref{fig:uvotspectrum}. Also shown (max-min) is the spectral shape derived from the difference between the maximum and minimum observed UVOT fluxes.  At $X(t)=-7.3$, the extrapolated UVW2 flux is zero, which provides a lower limit on the contribution of the host galaxy. Together with the fluxes of all the other UVOT bands extrapolated to $X(t)=-7.3$ we can derive the spectrum of this contribution, shown by blue stars ('avg') in Fig.~\ref{fig:uvotspectrum}. In Fig.~\ref{fig:uvotspectrum} we can indeed see that the variable (disc) component is blue and the constant (host galaxy) component is red.

\section{Comparison with Simple Accretion Disc Model}
\label{sec:disc}

To produce the impulse response functions shown in Fig.~\ref{fig:responses} we assumed an accretion disc as described by \cite{shakura73}, surround a black hole with the mass and accretion rate of NGC~4593 as listed in Table~\ref{tab:mmdot}. We assume an inclination of $45^{\circ}$. The choice of inclination does not have too great an effect on the median arrival time of reprocessed light but it has a very large effect on the time of the peak of the response. 
For an inclination of $45^{\circ}$, the peak of the response is a factor of $\sim3$ smaller than the median lag.

We assume a Schwarzschild black hole with an inner disc radius of 6~$R_{G}$. The exact value of the outer disc radius, assuming it is greater than a few hundred $R_{G}$, does not matter much. For a Kerr black hole with the disc again reaching to the innermost stable orbit, the median lag decreases by almost a factor 2. Thus, if all other disc parameters were well defined, lag measurement could, in principle, provide another method of black hole spin measurement, or at least of inner disc radius measurement.
We assume illumination by a point X-ray source located 6~$R_{G}$ above the axis of the black hole. We take the illuminating luminosity from the Swift BAT observations, extrapolating to 0.1-195 keV (Table~\ref{tab:mmdot}). The albedo is not well known. It is probably high near the inner edge of the disc where the surface of the disc may be ionised, but may be lower further out. Here we assume an albedo of 0.8 but assuming an albedo of 0.2 only increases the median lag by 9 per cent.

We also consider only the variable component of the UV/optical emission, produced by reprocessing of high energy emission, rather than the total emission from the disc which would include emission produced by dissipation of gravitational potential energy from accreting material. As the illuminating high energy emission heats the disc, the reprocessed variable emission is associated with larger radii than the same wavelength of emission from the quiescent disc.

To determine how we should translate the response functions into predicted model lags we simulated a UVW2 lightcurve with the observed X-ray lightcurve as input. Using the FR/RSS method the measured lag using the centroids of the correlation functions was $0.131 \pm 0.024$ and the lag using the peaks of the correlation functions was $0.076 \pm 0.03$. The model median lag from the response functions is 0.096, which corresponds reasonably to the measured lags. The arrival time of the peak of the response function is typically one third of the median lag and so is much shorter than the measured lag. We therefore take the median lag of the response functions as our model lag.

\subsection{Comparison of observed and model UVW2 lightcurves: Timescale dependence of lags}

\begin{figure}
\hspace*{-10mm}
\includegraphics[width=50mm,height=85mm,angle=270]{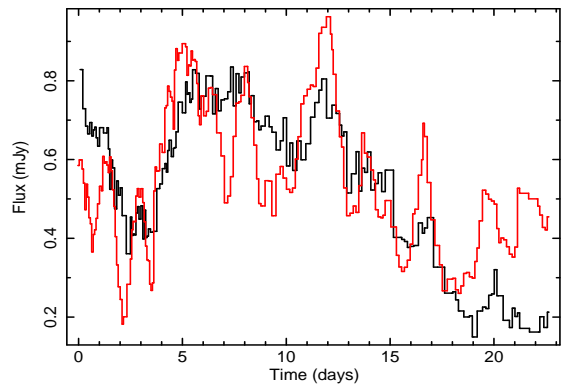}
\caption{Observed UVW2 lightcurve (black) with model UVW2 lightcurve (red) based on a simple disc reprocessing. The normalisation of the simulated lightcurve is arbitrary. The zero point is the start of the intensive monitoring period.}
\label{fig:w2sim}
\end{figure}

Using our disc model, with parameters given above, we can simulate the expected UVW2 lightcurve, assuming illumination of the disc by the observed X-ray lightcurve. The resulting model UVW2 lightcurve and the observed UVW2 lightcurve are shown in Fig.~\ref{fig:w2sim}. Fluxes are not yet included precisely in the simulation code and so the model lightcurve has been arbitrarily normalised to the same mean as the observed lightcurve. A `quiescent' level of 1~mjy, which is $\sim0.2$mJy below the lowest observed UVW2 flux was removed from the UVW2 lightcurve to highlight just the variable component.

The overall shapes of the two lightcurves are similar, with all of the large variations in the predicted model lightcurve having counterparts in the observed lightcurve, though the variations relative to the mean level are larger in the model lightcurve. As has been noted by a number of previous authors \cite[e.g.][]{berkley00,arevalo08_2251}, then for a lamp post X-ray model, it is necessary to have a large X-ray source height ($\sim100$~$R_{G}$) and a similarly large inner disc radius if the amplitude and degree of smoothness of the observed UV variability are to be reproduced, assuming that all of the reprocessing is carried out on an accretion disc.  We confirm that, if all of the observed UVW2 variability is to be produced by reprocessing of X-rays by just an accretion disc, then some blurring mechanism such as a large X-ray source height or scattering through the inner edge of the accretion disc \citep{gardner16} would be needed to reproduce the observed UVW2 amplitude of variability.

There are differences between the observed and model lightcurves in the long term variations, eg with the model lightcurve rising in the last 5 days of the observations whereas the observed lightcurve continues the decrease that has been going on for the previous $\sim15$ days. Although the peaks appear to line up reasonably well, {\sc JAVELIN} gives a lag of the model UVW2 by the observed UVW2 of $0.62 \pm 0.03$d, which is similar to the lag of the observed X-rays by the observed UVW2. The FR/RSS method gives a similar lag for the distribution of the centroids of the correlation functions of $0.55 \pm 0.10$d, although a shorter lag ($0.28 \pm 0.12 $d) for the distribution of the peaks of the correlation functions. The mean peak correlation coefficient is 0.74. 

\begin{figure}
\hspace*{-10mm}
\includegraphics[width=50mm,height=85mm,angle=270]{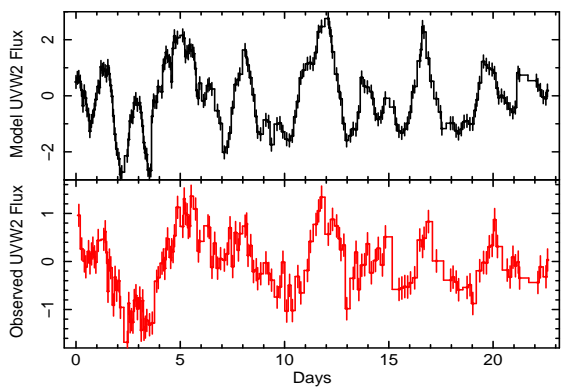}
\caption{The model (upper panel) and observed (lower panel) UVW2 lightcurves from Fig.~\ref{fig:w2sim} following subtraction of the mean level derived from smoothing with a 5d wide boxcar.}
\label{fig:boxcar}
\end{figure}

If we remove the mean level derived from smoothing with a boxcar, the similarity between the remaining observed and model lightcurves increases, eg in Fig.~\ref{fig:boxcar} where a boxcar of width 5d has been used. A similar relationship between the X-ray and UVW2 lightcurves was also seen in NGC~5548 after long timescale variations had been removed (Fig.5 of \cite{mch14}).
The similarity between the observed and model lightcurves indicates a strong causal relationship, as far as the short timescale variability is concerned, between the driving X-ray lightcurve and the observed UVW2 lightcurve. For the lightcurves shown in Fig.~\ref{fig:boxcar} the lag derived by {\sc JAVELIN} is $0.20 \pm 0.03$d and the FR/RSS mean centroid lag is to $0.13 \pm 0.05$d. The lags decrease as the boxcar width is reduced. 
With data of very high time resolution and with very short autocorrelation timescales it might be possible to distinguish separate peaks in crosscorrelation functions. However, with most real data, including the present data, the peaks are blurred together and so the effect of adding a long timescale lag to a short timescale lag is actually to produce a correlation function with an intermediate lag. 

These results show that there are longer timescale variations in the observed UVW2 lightcurve which are not reproduced by modelling by an accretion disc. However the short timescale ($<5$d) variations can, at least to first order, be reproduced by reprocessing of the X-rays, or by some other high energy emission with similar variability properties to the X-rays, by an accretion disc. There is still some additional lag of the observed UVW2 over and above that expected from reprocessing, but it is small, and decreases as lower frequency variations are filtered out. We note that the observed lag of the unfiltered X-rays by UVW2 is $\sim 0.66$d. This lag is much longer than might be explained by an X-ray source of size $\sim100$~$R_{G}$ or a similarly large inner disc radius as, for NGC~4593, $\sim100$~$R_{G}$ corresponds to 0.04d. 

If the inflated inner edge of the accretion disc acts as a scatterer for the hard X-rays, adding an addition lag as in the model of \cite{gardner17}, the fact that the X-ray to UVW2 lag becomes smaller as we remove long timescale variations may indicate a stratified scatterer. Thus X-rays which scatter from the top of the inflated inner disc may only undergo a small number of scatterings. They would therefore not suffer a large additional delay beyond the light travel time expected from direct illumination of the outer disc. Similarly the reprocessed signal in the UVOT bands should not be greatly blurred. However X-rays which travel through a greater path length of scatterer, along the mid plane of the disc, will suffer longer delays and will also result in a more blurred reprocessing signal. In this model the total UVOT lightcurves should be a sum of all of the reprocessed contributions.

\subsection{Energetics}

We can make a rough determination of whether the X-rays could be directly powering the UV/optical variations by comparing the X-ray luminosity hitting the disc, for various source geometries, and the luminosity in the varying (not the steady) component of the UV/optical fluxes.  The BAT 70 month survey gives an average 14-195 keV luminosity of $1.58 \times 10^{43}$ ergs s$^{-1}$. Extrapolating to 0.1-195 keV with $\Gamma=1.75$ gives $3 \times 10^{43}$ ergs s$^{-1}$ which we take as an approximate estimate of the total X-ray luminosity. As the lowest observed X-ray count rates are almost zero, we assume that all of the X-ray emission comes from the central varying compact source and that there is no significant quiescent component.

\begin{figure}
\hspace*{-10mm}
\includegraphics[width=50mm,height=85mm,angle=270]{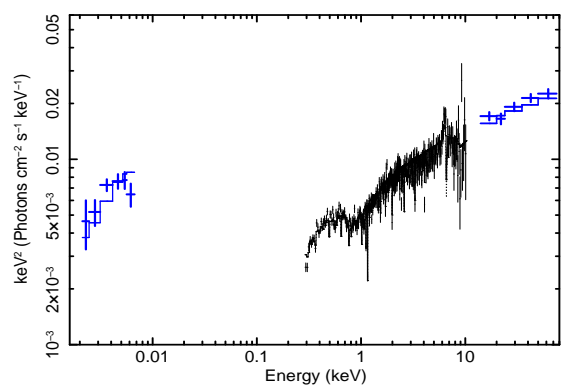}
\caption{Broad band $\nu F(\nu)$ SED of NGC~4593, including the variable UVOT components, fitted to the same model as in Fig.~\ref{fig:xspec}, together with a disk black body in the UVOT bands.}
\label{fig:sed}
\end{figure}

To estimate the total varying luminosity over the UVOT bands we have fitted models in {\sc xspec} to the data shown in Fig.~\ref{fig:uvotspectrum}. The models are not well constrained
but the total flux, $\sim 1.16 \times 10^{-11}$ \ecs over the range covered by the UVOT ($\sim 1850-5500 \AA$), does not vary by more than 10\% between models giving, for a distance of 35 Mpc, a luminosity of $\sim 1.7 \times 10^{42}$ ergs s$^{-1}$.

To estimate the solid angle subtended at the X-ray source by the UV/optical emitting regions of the disc, we estimate, from Fig.~\ref{fig:responses}, that the bulk of the UV and optical emission detected by the UVOT comes from between radii of 10-300 $R_{G}$ (1 $R_{G}$ = 38s).
For an X-ray point source at a height, $H$, of 6~$R_{G}$, this range of the disc subtends 3.22 sr, ie a covering fraction of 0.257 (or 0.346 for $H=10$ $R_{G}$). Conservatively taking the lower figure gives an X-ray luminosity of $7.7 \times 10^{42}$ ergs s$^{-1}$ impacting the UVOT-emitting part of the disc, ie about $4.5 \times$ the observed UVOT luminosity. Thus even with a disc albedo of $\sim80$ per cent, there would be sufficient X-ray illumination to power the observed UV/optical emission. 

As an alternative way of demonstrating the energetics we show, in Fig.~\ref{fig:sed}
the broad band SED.  The exact model fit in the UVOT bands is not important but we can see that the total luminosity in the UVOT bands does not exceed that in one decade of the X-ray spectrum.

There are, of course, a number of uncertainties in this calculation, eg the albedo, the X-ray source size, the illuminating X-ray energy band and the inner radius of the emission region (the outer is relatively unimportant). We may also wish to consider the emission at shorter wavelengths than UVW2, but then we would have to also decrease the inner radius which would increase the solid angle. Here we therefore conclude that, at least to first order, there is sufficient X-ray luminosity to power the observed UV/optical variations.

The above accretion disc modelling is quite basic, but serves to show that although reprocessing by an accretion disc can reproduce a good deal of the short timescale UV variability, there are variations on longer timescales, with larger lags, which cannot be explained by reprocessing purely by an accretion disc. To explain all of the UV/optical variations by reprocessing of X-rays, we therefore require more than just an accretion disc as the reprocessor.
Below we provide more sophisticated modelling to derive the reprocessing functions needed to reproduce all variations.

\begin{figure*}
\begin{tabular}{c}
\includegraphics[width=16cm,height=21cm]{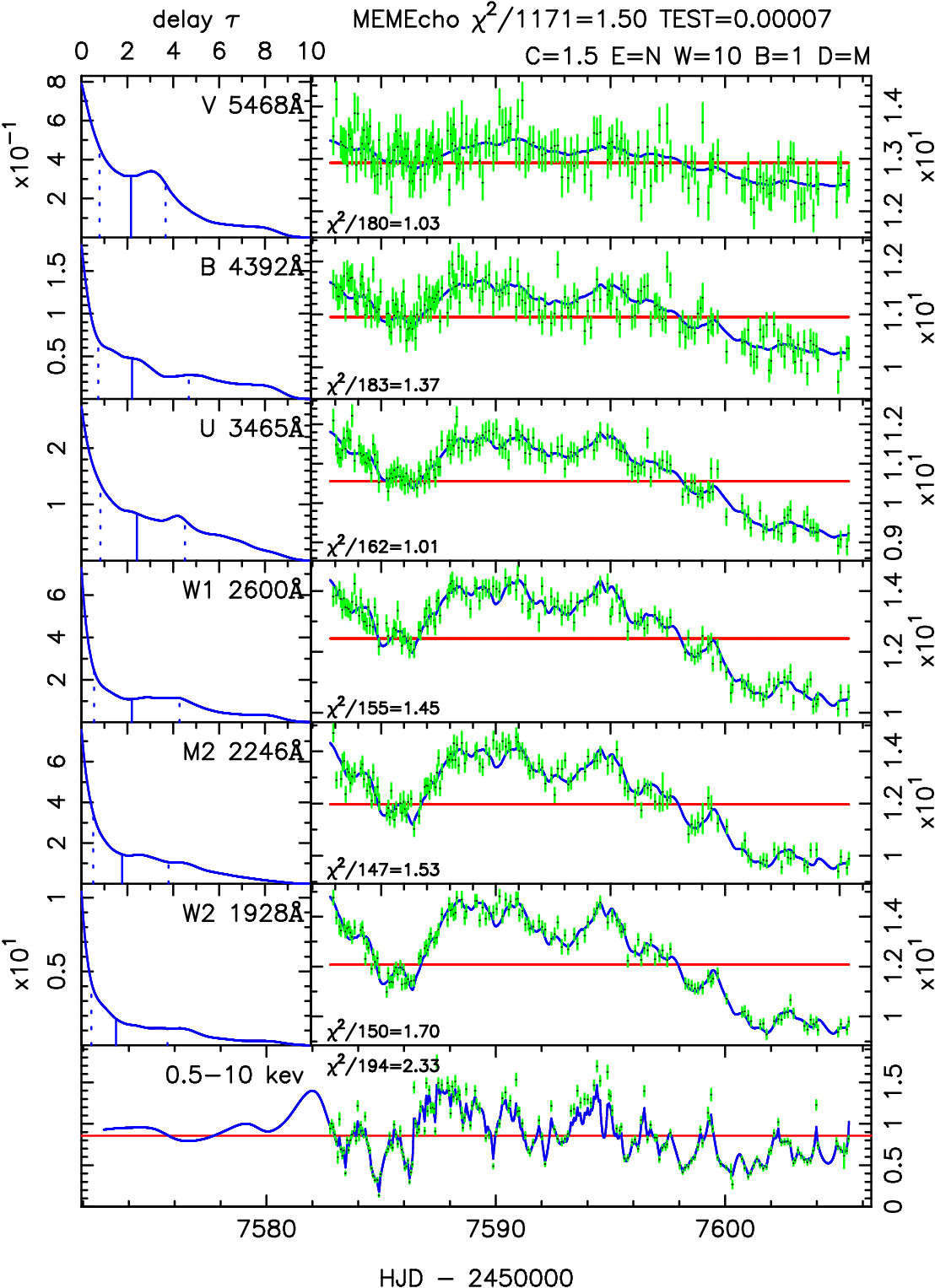}
\end{tabular}
\caption[] {\small MEMEcho fit of the linearised echo model to Swift lightcurves of NGC~4593. {\it Bottom panel:} X-ray (0.5-10~kev) lightcurve data along with the fitted model $X(t)$. {\it Left column:} Delay distributions $\Psi(\tau|\lambda)$ for three UV (UVW2, UVM2, UVW1) and three optical (U, B, V) bandpasses. {\it Right column:} Swift lightcurve data along with the echo lightcurves obtained by convolving $X(t)$ with $\Psi(\tau|\lambda)$. Vertical (blue) lines indicate the median (solid) and quartiles (dashed) of the delay distributions. Horizontal (red) lines indicate the reference levels $X_0$ for the X-ray lightcurve and $F_0(\lambda)$ for the echo lightcurves. The fit shown is for $\chi^2/N=1.5$ with $W=10$ to make $X(t)$ 10 times more flexible than $\Psi(\tau|\lambda)$. Note the two-component structure of the delay maps, with a prompt response peak at $\tau=0$ and extended response reaching to $\tau=10$~d.
\label{fig:memecho}
}
\end{figure*}
\typeout{fig:memecho}

\section{MEMEcho modelling of X-rays driving the UV and optical variations}
\label{sec:memecho}

Visual inspection of the lightcurves and cross-correlation analysis indicates that
a physical relationship exists between the X-ray variations and the UV/optical variations. In order to test the reality of this relationship, and to infer delay distributions rather than just mean lags for each of the UV/optical lightcurves, we here fit the lightcurves in detail, under the assumption that the X-ray variations are driving time-delayed responses in other bands. 

We employed the maximum entropy fitting code MEMEcho \citep{horne94,horne04}
to fit the Swift lightcurve data with a {\it linearised} echo model
\begin{equation}
        F(t|\lambda) = F_0( \lambda ) + \int_0^{\tau_{\rm max}}
        \Psi( \tau | \lambda )\, \left( X(t-\tau)-X_0 \right)\, d\tau
\ .
\end{equation}
Here $X(t)$ is the driving (X-ray) lightcurve. $X_0$ is an arbitrary
reference level, set at the median of the X-ray data.
$F_0(\lambda)$ is the corresponding reference level 
in the echo lightcurve and $\Psi(\tau | \lambda)$ is the
delay map (i.e. the response function) for the echo 
response at wavelength $\lambda$.

Fig.~\ref{fig:memecho} shows a MEMEcho fit to the Swift lightcurve data, where $X(t)$ is adjusted to fit the X-ray (0.5-10 keV) lightcurve, and the reference levels $F_0(\lambda)$ and delay distributions $\Psi(\tau|\lambda)$ are adjusted to fit the three UV lightcurves (UVW2, UVM2, UVW1) and the three optical lightcurves (U, B, V).

MEMEcho adjusts the model X-ray lightcurve and the echo reference levels and delay maps to find a ``good" fit to all observed lightcurves, i.e. X-ray, UV and optical, with a ``simple" model. The ``good" fit requirement is imposed on the model by requiring a specific value of $\chi^2/N$, where $N=1171$ is the number of data values. The competing requirement of a ``simple" model is achieved by maximising the entropy $S$ of the functions $X(t)$ and $\Psi(\tau|\lambda)$.

We sample $X(t)$ and $\Psi(\tau|\lambda)$ on uniform grids in $t$ and $\tau$, with the same fine spacing $\Delta t=0.01$~d in order to resolve the most rapid structure in the X-ray lightcurve. We restrict the delay map to $\tau$ in the range $0$ to $10$~d, with $\tau>0$ to impose causality (no UV/optical response prior to X-ray variations) and $\tau<10$~d since this is roughly 1/3 of the timespan of the data and there is no sign, in the simple cross-correlations above of any strong correlation on timescales greater than $\sim2$d.

The entropy of the discretely sampled positive 
function, $p(t)=p_i$ at $t=t_i$, is defined as
\begin{equation}
        S(p) = p_i - q_i - p_i\, \ln{p_i/q_i}
\ ,
\end{equation}
where for each $p_i$ the default value $q_i$ is the geometric mean of its neighbors, $q_i=\left(p_{i-1}\,p_{i+1}\right)^{1/2}$. With this definition the entropy has a regularising effect, expressing a preference for smooth positive functions, favouring Gaussian peaks and exponential tails. A parameter $W$ scales the entropy weight of the the echo delay maps relative to that of the X-ray lightcurve, controlling their relative flexibility.

In fitting the Swift lightcurves, we ran MEMEcho repeatedly to find a grid of models fitting the data with $\chi^2/N$ ranging from 2 to 0.8, and for $W=1$ and $10$. Here the $\chi^2$ level controls the trade-off between resolution and noise in the resulting delay maps. For higher $\chi^2$ the model lightcurves are too smooth to follow significant features in the lightcurve data. For lower $\chi^2$ the model fits more subtle features in the lightcurve data, but to do so the delay maps, and to a lesser extent the model X-ray lightcurve, develop larger-amplitude fine-scale structure that eventually, for the smallest $\chi^2$, becomes unacceptable.

Fig.~\ref{fig:memecho} presents the MEMEcho fit for $\chi^2/N=1.5$ and $W=10$, which we judge by eye to achieve the right balance, giving a ``good" fit that is not excessively ``noisy". Roughly speaking $W=10$ allows the X-ray lightcurve $X(t)$ to be 10 times more flexible than the delay maps $\Psi(\tau|\lambda)$. This is appropriate since the most rapid structure in the X-ray lightcurve is not evident in the echo lightcurves.

Notice that the delay maps share a similar structure, with a prompt response peak at $\tau=0$, roughly 1/4 of the response at $\tau<1$~d, plus a broader response tail decreasing toward $\tau=10$~d. The extended response is generally stronger at longer wavelengths. Small bumps and wiggles on the extended response do not show a clear trend with wavelength. These may be unreliable artefacts arising from attempts to fit the echo model to residual systematic errors in the lightcurve data and/or to variations that are not driven by the X-ray variations.

The fitted model adequately represents most of the features in the lightcurve data, consistent with the X-ray variations driving linear responses in the UV and optical light. Evidence for the prompt response peak at $0<\tau<1$~d arises from many X-ray features with well-detected counterparts in the echo lightcurves. For example, the three X-ray dips during 7583-85 have corresponding dips in the UV echo lightcurves, as do the peaks at 7599 and 7602. The more extended response is needed to produce the gradual decrease in the echo lightcurves.

A few minor defects in the fit may also be noted. The 1-d X-ray dip at 7590 has no counterpart in the echo lightcurves. The 2-day peak near 7494 in the echo lightcurves is well modelled, but the corresponding X-ray feature is stronger than required, so that the model X-ray lightcurve falls below the X-ray data in this region. In particular, the fit inserts a narrow X-ray dip in a data gap near 7594.6.

To summarise, our main conclusion is that the linearised echo model provides an acceptable fit to the Swift lightcurve data in all wavebands. We require that the delay structure has a prompt response peak with $0<\tau<1$~d to produce rapid correlated variations on 1-day timescales while washing out, from the UV/optical lightcurves, sub-day structure seen in the X-ray lightcurve. We also require an extended tail to the delay structures to produce a gradual decline in the UV/optical echo lightcurves.
The rapid response would naturally be produced by reprocessing on a nearby accretion disc with the more extended response being produced in the broad line region gas. This latter possibility is discussed in more detail in \cite{cackett17}.

\section{Discussion}
\label{sec:discussion}

\subsection{Comparison of AGN Lag Spectra: Cosmological standard measuring rods?}

\subsubsection{UVW2 to V-band lags}

Well measured X-ray/UV/optical lags as measured by Swift have already been published for NGC~4151 \citep{edelson17_4151} and NGC~5548 \citep{mch14, edelson15} and well measured lags between the X-ray and UVW1 from XMM-Newton and ground based g-bands in NGC~4395 have also been published \citep{mch16}. These observed lags, together with the lags derived here from NGC~4593 and
model predictions assuming reprocessing on an accretion disc, using the model code from which the response functions given above (Fig.~\ref{fig:responses}), are given in Table~\ref{tab:mmdot}. In Fig.~\ref{fig:w2vlags} we present the ratio of the measured to model lags for the lag of the UVW2 band by the V-band. One might measure the lag between any two bands from a fit to the overall lag spectrum, which would produce a smaller uncertainty. However such fits can involve assumptions, eg about which bands to include, and so here we use the simple observed lag between the UVW2 and V bands.
In Fig.~\ref{fig:xw2lags} we present the ratio of the measured to model lags for the lag of the X-rays by the UVW2 band. 
The model lags are based on the best current measurements of the black hole masses and accretion rates. Although not plotted here as its black hole mass is not as precisely known, the observations of NGC~2617 \citep{shappee14}, which were the first to point out the possible discrepant lag between the X-ray and UV/optical bands, are in broad agreement with Figs.~\ref{fig:xw2lags} and \ref{fig:w2vlags}.

We can see that the UVW2 to V-band lags are a factor $\ltsim 2$
of our present predictions from a standard accretion disc \citep{shakura73}. 
The ratio for NGC~4395 is very close to unity. In this case, however, the X-ray and UV lightcurves were from XMM-Newton and do not therefore sample long timescales and we have shown above that if longer timescale variations are included, longer lags are measured. By filtering out long timescale variations, lags more consistent with disc reprocessing may be obtained. 

The discrepancy of $\ltsim 2$ is broadly consistent with the factor of 2.25 which 
we measured in NGC~5548 in \cite{mch14} and with the factor $\sim3$
which was quoted by subsequent reverberation mapping programmes \cite[e.g.][]{edelson15,fausnaugh16,edelson17_4151}. Microlensing observations \cite[e.g.][]{morgan10,dai10,mosquera13} typically quote a somewhat larger discrepancy, $\sim 4 \times$, but the methods used to measure the discrepancy are quite different.

We note that the deduced disc size discrepancy may appear slightly different from one set of observations to another. For example, in NGC~5548 the UVW2 to V-band lag from \cite{mch14} is $1.35^{+0.33}_{-0.35}$d but from \cite{edelson15} is it $1.16 \pm 0.53$d, although these lags are perfectly consistent within the errors. We also note that the model lags used here are based on the time for half of the reprocessed light to appear, as derived from a numerical computer model of the disc, rather than on an analytical calculation, which may have different assumptions, eg the value of the $'X'$ parameter in equation 9 of \cite{fausnaugh16}.
However, perhaps most importantly, the wavelength range over which the comparison between model and observation is carried out can have a big effect on the deduced discrepancy \cite[e.g. see][]{fausnaugh16} as the observed lags in the different wavebands do not generally all lie smoothly on the same curve. The accompanying paper based on HST observations \cite{cackett17} includes lags from both shorter and longer wavelengths than those used here and the inclusion of those additional lags leads to almost twice as large a discrepancy, assuming the analytic model predictions of \cite{fausnaugh16}, than if just the Swift lags are used.

\subsubsection{X-ray to UVW2 lags}

The ratios of observed to model lags for the X-ray to UVW2 bands are, with the exception of NGC~4395, far from the expectations of a standard accretion disc and also vary considerably from one AGN to another. It is not clear why the ratio for NGC~4395 is very close to unity. The lack of long timescale variations in the XMM-Newton lightcurves have already been noted, but NGC~4395 also has by far the lowest accretion rate of any AGN considered here. It is therefore possible that the inner edge of the accretion disc is less inflated than in higher accretion rate AGN, thereby providing less of a scattering impediment to X-ray illumination of the outer disc. 
The largest ratio is for NGC~4151, which is also the most absorbed AGN in this small sample. This observation is in agreement with the hypothesis that absorption and scattering of the X-rays affects the lag of the UV/optical bands \citep{gardner16}.

The above results indicate that the UVW2 to V-band lags might be useful as a cosmological standard measuring rod \citep[c.f.][]{cackett07} but any lags relative to the X-ray band must be treated with great care.

\subsection{Geometry of the Inner Region: An extended reprocessor}

The lags measured from unfiltered lightcurves show, in all cases, that the lags of all of the other UV/optical bands (except U-band) relative to UVW2 are close to the predictions from the accretion disc model of \cite{shakura73} but the lag of the X-rays by the UVW2 band are much larger than predicted.
\cite{gardner16} propose that the X-rays do not directly illuminate the outer accretion disc but instead heat up the inner accretion disc which re-radiates onto the disc in the far-UV. The additional lag between the X-ray and UVW2 bands is then the thermal timescale for the X-rays to heat the inner disc. There are a number of attractive aspects of this model. The large far-UV emission region helps to smooth out the rapid variations in the X-ray band which would otherwise predict more rapid variations in the UV and optical bands than are observed. If the inner disc is very inflated by the X-ray heating, then the additional height of the emission region increases the solid angle subtended by the outer disc, thereby ensuring that there is adequate luminosity impacting on the outer disc to account for the observed UV and optical variations. A large far-UV emission region might be expected to have a tail of emission extending into the soft X-ray band. A weak soft excess has been seen by previous observers \citep[e.g.][]{brenneman07_4593} but we do not find such a component in the present Swift data.

Here we have shown here that when the X-ray and UV lightcurves are filtered to remove long timescale variations, the X-ray to UVW2 lag becomes close to the predictions of disc reprocessing. A similar conclusion was reached regarding the lags in NGC~5548 by \cite{mch14} and, for NGC~4395 whose lags were measured using XMM-Newton \citep{mch16} and therefore long timescale variations were not measured, the lag ratio is very close to the expectations of disc reprocessing. These results show that although short timescale UV/optical variability is consistent with X-ray reprocessing on a disc, there are longer timescale variations in the UV and optical wavebands than cannot be accounted for by simple reprocessing of X-rays by an accretion disc. With MEMEcho analysis we have shown here that the X-rays can be a reasonable driver of the unfiltered variations in all bands as long as we allow for a complex resprocessor where, in addition to a close reprocessor such as the accretion disc, the reprocessing material extends to distances of a few lightdays from the illuminating source. The broad line region gas thus provides a natural explanation for the extended reprocessing region. This conclusion is consistent with the large excess lag seen in the U-band which is almost certainly Balmer continuum emission resulting from reprocessing of high energy emission in the BLR. We incidentally note that the excess U-band lag in NGC~4593 is larger than in the lower accretion rate AGN NGC~5548 and NGC~4151 where the inner edge of the BLR will be closer to the black hole. Although the data at present do not allow detailed analysis 
it might be interesting in future to see whether the U-band excess lag does vary with accretion rate.

The geometry of the BLR is not well known but it is likely to subtend a large solid angle at the X-ray source. Thus worries concerning whether sufficient X-ray illumination hits the reprocessing to account for the resultant UV/optical emission are diminished. If we consider only the variable part of the UV/optical emission, the simple energetic arguments outlined above indicate that, although there is no great excess, there is probably enough X-ray illumination of the accretion disc to account for the observed short timescale variability.\\

\section{Conclusion}

Although \cite{starkey17} find that the multi-band observations of NGC~5548 cannot be satisfactorily accounted for simply by reprocessing of X-rays, the Swift UV/optical lightcurves of NGC~4593 can be reasonably reproduced by reprocessing of X-rays, but by a complex reprocessor including a nearby accretion disc and more distant broad line gas clouds. 

The observation of a large additional lag between the X-ray and UVW2 bands in all well studied AGN, above the expectation of simple reprocessing from a nearby accretion disc, is at first sight good evidence for scattering of X-rays in the inner accretion disc. However the measured additional lag becomes closer to the prediction of direct disc reprocessing both here in NGC~4593 and also in NGC~5548 \citep{mch14} as one filters out long timescale variations, which may come from the broad line region gas. In the case of multiple or extended reprocessors then, in perfect data, we would be able to detect separate peaks corresponding to different lags in the cross correlation function. However, with the present data, the effect is to produce asymmetric correlation functions where the peak is some combination of the individual contributing lags. Thus reprocessing in the BLR may explain at least part of the additional X-ray/UVW2 lag.

It is not known how the UVOT spectrum extrapolates to shorter wavelengths. Thus we do not know how much reprocessed emission we need to produce. However there appears to be enough energy in the X-ray band to account, by direct illumination, for the variable part of the UVOT emission, assuming a ''standard'' X-ray source of size 10$R_{G}$. 

There remains slight uncertainty regarding the amplitude of
variability of the reprocessed UVW2 emission. Simple reprocessing of
the observed X-ray lightcurve only from an accretion disc produces
larger amplitude UVW2 variability than is observed. However MEMEcho
analysis of reprocessing from an extended reprocessor can reproduce a
model UVW2 lightcurve which is an acceptable fit to the observed
lightcurve. The only caveat here is that the input X-ray lightcurve is
not the directly observed lightcurve but is a fit to the X-ray
lightcurve. However it is a reasonable fit, ie a reasonable
description of the underlying lightcurve, given the noise involved in all observational measurements. 

For the small number of AGN monitored so far, the UV/optical lags scale broadly similarly with mass and accretion rate, typically being a factor $\ltsim$2 of the expectations from disc reprocessing (for the UVW2 to V-band lag). The X-ray to UVW2 lags, however, are all larger than the expectations from disc reprocessing, and the discrepancy varyies between AGN, with the most absorbed AGN monitored so far, NGC~4151, showing the largest discrepancy. This res
ult indicates a probable role for scattering in producing the X-ray to UV lags in at least some AGN.

\begin{table*}
\begin{tabular}{lllllcccc}
  AGN   & $M$   & $L_{Bol}$&\me &$L_{ion}$& Model       & Model      &  Observed     & Observed  \\
        & $10^{6}$\msun & ergs/s & \% & ergs/s & X-ray - UVW2 & UVW2 - V   &  X-ray - UVW2 &  UVW2-V \\
        &       &        &    &        & lag (days)   & lag (days) & lag (days)    & lag (days)  \\
&&&&&&&&\\
NGC~4151 & $37.6 \pm 11.5$&$1.0 \pm 0.5 \times 10^{44}$ &2.1 & $2.6 \times 10^{43}$ & 0.174  & 0.425 & $3.58 ^{+0.36}_{-0.46}$ & $0.96^{+0.47}_{-0.46}$ \\ 
NGC~4395 & $0.36 \pm 0.11$&$5.4 \times 10^{40}$         &0.12& $1.6 \times 10^{40}$ &$0.00293$ &$0.00748$ & $4.05^{+0.54}_{-1.13} \times 10^{-3}$ & $5.07^{+0.74}_{-1.29} \times 10^{-3}$\\ 
NGC~4593 & $7.63 \pm 0.16
$&$7.8 \pm 3.5 \times 10^{43}$ &8.1 & $3 \times 10^{43}$   & 0.090  & 0.231 & $0.66 \pm 0.15$ & $ 0.247 \pm 0.38$   \\ 
NGC~5548 & $52.3 \pm 1.9 $&$3.2 \pm 2.0 \times 10^{44}$ &4.8 & $1.0 \times 10^{44}$ & 0.286  & 0.698  & $1.12 \pm 0.49$ & $1.16 \pm 0.53$    \\ 
\end{tabular}
\caption{Black hole masses, accretion rates and resultant accretion disc model lags between the X-ray and UVW2 and between the UVW2 and V-bands. 
For NGC~4395 both the mass and the bolometric luminosity are from \protect\cite{peterson05}. The masses for NGC~4151, NGC~4593 and NGC~5548 are taken from \protect\cite{bentz15} and the bolometric luminosities used in the derivation of the accretion rates are the mean and spread of observations reported by \protect\cite{vasudevan09a,vasudevan09b,vasudevan10,woo02a}. 
The ionising luminosity is taken from the BAT 70 month survey, extrapolated to 0.1-195 keV.
Here we assume an X-ray source height of $6 R_{g}$, an inclination of $45^{\circ}$, an albedo of 0.8 and an inner disc radius of $6 R_{g}$. The observed lags for NGC~4593 are from \protect\cite{mch16}, for NGC~4151 from \protect\cite{edelson17_4151}, for NGC~5548 from \protect\cite{edelson15} and the results for NGC~4593 are from the present work. 
In all cases the lags are derived from the centroid distributions using the FR/RSS method and no long term variations have been removed from any lightcurve. For NGC~4395 where the observed UV band was UVW1, the lag has been corrected to the UVW2 band assuming a scaling with wavelength to a power 1.15, which is intermediate between the expected disc value of 1.333 and the value obtained by observation which, roughly, is close to unity. For NGC~4151 and NGC~5548, X-ray lags were derived from the X4 band. For NGC~4593 where there is no evidence of variation of lag with X-ray energy, the full 0.5-10 keV band is used, as it is for NGC~4395 where the source is too faint to measure lags relative to different energy bands.
}
\label{tab:mmdot}
\end{table*}

\begin{figure}
\hspace*{-5mm}
\includegraphics[width=60mm,height=80mm,angle=270]{fig16_v_model_data.ps}
\caption{The ratio of observed UVW2-V band lags compared to the predictions from reprocessing on a standard optically thick accretion disc \protect\cite{shakura73}. The data from which this plot is produced are given in Table~\ref{tab:mmdot}.
 }
\label{fig:w2vlags}
\end{figure}

\begin{figure}
\hspace*{-5mm}
\includegraphics[width=60mm,height=80mm,angle=270]{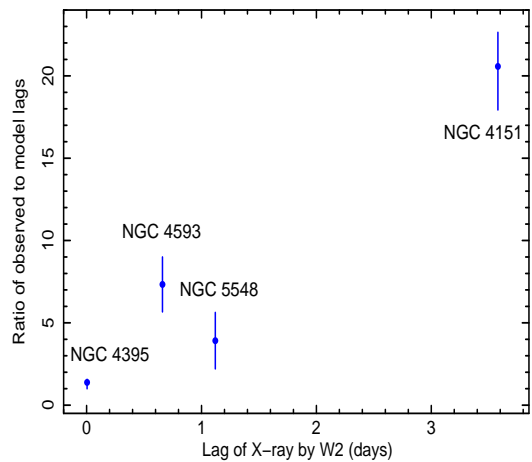}
\caption{The ratio of observed X-ray to UVW2 band lags compared to the predictions from reprocessing on a standard optically thick accretion disc \protect\cite{shakura73}. The data from which this plot is produced are given in Table~\ref{tab:mmdot}.
 }
\label{fig:xw2lags}
\end{figure}

\section*{Acknowledgements}
This paper is dedicated to the memory of Neil Gehrels. Without his inspired leadership of the Swift mission the research presented here, and in other similar AGN monitoring programmes, would not have been possible. IMcH acknowledges support from a Royal Society Leverhulme Trust Research Fellowship LT160006 and from STFC grant ST/M001326/1. KH acknowledges support from STFC grant ST/M001296/1 and JG acknowledges support from NASA under awards NNX15AH49G and 80NSSC17K0126.
This work made use of data supplied by the UK Swift Science Data Centre at the University of Leicester. 

\input{references.bbl}

\end{document}